\documentclass[a4paper,12pt]{iopart}
\usepackage{color}
\usepackage{amsmath}
\usepackage[T1]{fontenc}
\usepackage[utf8]{inputenc}
\usepackage{graphicx}
\usepackage{textcomp}
\usepackage{bm}
\usepackage{iopams}
\usepackage{dsfont}
\usepackage{mathrsfs}
\usepackage{fixmath}
\usepackage{exscale}
\usepackage[colorlinks=false, pdfborder={0 0 0}]{hyperref}
\usepackage{letltxmacro}
\LetLtxMacro{\originaleqref}{\eqref}
\renewcommand{\eqref}{Eq.~\originaleqref}
\newcommand{\figref}[1]{Fig.~\ref{#1}}
\newcommand{\imag}{\rmi}

\newcommand{\dd}{\mathrm{d}} 
\newcommand{\pdd}{\mathrm{\partial}} 

\newcommand{\pdiff}[3][]{\frac{\pdd^{#1} #2}{\pdd #3^{#1}}}

\newcommand{\qty}[1]{\left(#1\right)}
\newcommand{\sqty}[1]{\left[#1\right]}

\newcommand{\abs}[1]{\left|#1\right|}

\usepackage{dcolumn}
\usepackage{bm}
\usepackage{multirow}

\begin{document}
\title{Matter-wave lensing of shell-shaped Bose-Einstein condensates}

\author{Patrick Boegel$\mathrm{^{1}}$, Alexander Wolf$\,\mathrm{^{2}}$, Matthias Meister$\mathrm{^{2}}$, and Maxim A. Efremov$\mathrm{^{1,2}}$}

\address{$\mathrm{^1}$Institut f\"ur Quantenphysik and Center for Integrated Quantum Science and Technology (IQ$^{\rm ST}$), Universit\"at Ulm, D-89081 Ulm, Germany}
\address{$\mathrm{^2}$German Aerospace Center (DLR), Institute of Quantum Technologies, D-89081 Ulm, Germany}
	
\ead{patrick.boegel@uni-ulm.de}

\vspace{10pt}
	
\begin{abstract}
Motivated by the recent experimental realization of ultracold quantum gases in shell topology, we propose a straightforward implementation of matter-wave lensing techniques for shell-shaped Bose-Einstein condensates. This approach allows to significantly extend the observation time of the condensate shell during its free expansion and enables the study of novel quantum many-body effects on curved geometries. With both analytical and numerical methods we derive optimal parameters for realistic lensing schemes to conserve the shell shape of the condensate for times up to hundreds of milliseconds.
\end{abstract}


\vspace{2cm}
Version: \today

\maketitle

\section{Introduction}
\label{sec:Introduction}
Many-body physics on shell topology has recently experienced a huge progress with the first creation of shell-shaped quantum gases~\cite{Carollo.2022,Jia.2022}. Inspired by the near-term availability of this novel topology, theoretical research has lead to new insights into quantum phenomena such as vortices~\cite{Turner.2010,Padavic.2020,Bereta.2021}, the Berezinskii-Kosterlitz-Thouless transition~\cite{Kosterlitz.2016,Tononi.2021}, and the impact of dimensional as well as topological crossovers on the excitation spectrum~\cite{Padavic.2017,Sun.2018}. Consequences of the system being forced on a curved manifold may also influence ultracold chemistry~\cite{Perez-Rios.2020} and few-body physics \cite{Naidon_2017,Blume_2012}, for example, in the form of confinement induced resonances \cite{Olshani_PRL1998,DUNJKO2011461}.

Experimentally, two schemes have currently shown to be capable of creating shells: (i) radio-frequency (rf) dressing~\cite{Zobay.2001,Zobay.2004,Garraway.2016,Perrin.2017} in combination with a microgravity environment~\cite{Carollo.2022,Lundblad.2019}, and (ii) an optically confined mixture of two Bose-Einstein condensates (BECs) employing a magic laser wavelength for trapping the mixture~\cite{Wolf.2022,Meister2022,Jia.2022}. The latter scheme could also be realized in microgravity~\cite{Wolf.2022} and typically requires an additional magnetic Feshbach field~\cite{Timmermans.1999,Chin.2010} to tune the interspecies interaction that builds up the shell. 
Although it is quite the accomplishment that both schemes can produce shells of quantum gases, they require multiple electromagnetic fields being superimposed to keep the system confined in a stable ground state. Probing the shell in a clean and undisturbed environment to study any of the aforementioned quantum phenomena is therefore a challenge. By completely switching off the trap the shell could be probed during the subsequent free expansion. However, without a confinement, the matter-wave dynamics of the released shell generally leads to a rapid increase of its thickness due to the atoms moving both outwards and inwards. Thus, observing shell related physics would usually be limited to few tens of milliseconds, until the atoms have reached the center and the shell structure is lost~\cite{Lannert.2007,Tononi.2020,Wolf.2022,Jia.2022}.

In this article we explore the application of matter-wave lensing techniques~\cite{Chu1986,Ammann1997,Kovachy2015,Pandey_2021,Deppner_2021} to shell-shaped BECs in order to provide a better testbed for interesting physical phenomena. By preparing a free expansion in which the shell structure is conserved for hundreds of milliseconds, there is sufficient time to manipulate the shell by exciting vortices, exploiting Feshbach resonances, or performing a quench and observe the response of the system afterwards. Here we propose two schemes with their applicability depending on the setup initially used to prepare the shell: (i) delta-kick collimation (DKC)~\cite{Ammann1997,Kovachy2015,Pandey_2021,Deppner_2021}, i.e. a typical thin lens approach to reduce the width of the momentum space distribution at the expense of a larger position space distribution which is well suited for optically confined mixtures, and (ii) excitation induced collimation~\cite{Chu1986,Deppner_2021}, where a change of the trap parameters is used to induce an oscillation in the thickness of the shell before switching off the trap close to the point of maximum thickness. The latter is specifically designed to be used in setups based on rf dressing that do not allow the trapping potential to be switched completely off and on again at will due to mixing of different magnetic hyperfine sublevels. By deriving optimal conditions for the collimation sequence, we show that both techniques are capable of conserving the shell structure for several hundreds of milliseconds which can be used for extensive probing of the system.

This article is structured as follows. In Sec.~\ref{sec:section2} we introduce an effective model to describe both the ground state and dynamics of a large spherically symmetric shell-shaped BEC. Starting from the three-dimensional (3D) nonlinear Gross-Pitaevskii equation (GPE), we discuss two approximations to reduce the equation of motion to a one-dimensional (1D) linear Schrödinger equation and present a general analytic solution which is employed throughout the article. 
The DKC approach for lensing of shell-shaped BECs is presented in Sec.~\ref{sec:section3}, where we focus on controlling the thickness of the shell during its free expansion by varying multiple parameters involved in the lensing process. Using our effective model, we derive conditions to conserve the shell structure for the longest possible times. At the end of the section a generalized version of DKC is discussed that can be used to describe excitation induced collimation.
In Sec.~\ref{sec:section4} we consider the application of an adapted scheme to rf-dressed potentials, where any change of the parameters needs to be reasonably adiabatic in order to not mix different magnetic hyperfine sublevels. 
Our conclusion in Sec.~\ref{sec:section5} is followed by two appendices which present the effective 1D model in detail. \ref{Appendix:from_3D_GPE_to_1D_SE} contains an in-depth discussion concerning the approximations needed to apply the 1D model instead of the 3D GPE, whereas \ref{Appendix:effective_1D_model} summarizes the analytical solutions derived for different time-profiles of the lensing schemes.

\section{Shell-shaped Bose-Einstein condensates}
\label{sec:section2}

In this section we first present the general framework to study shell-shaped BECs with the 3D GPE and then introduce an analytical model based on the 1D Schr\"odinger equation, that enables an efficient description of 
the ground state and dynamics of shell-shaped BECs valid in the limit of large shell radius.

\subsection{System under study}

To describe a BEC consisting of $N$ atoms of mass $m$, in an external potential $V(\mathbf{r},t)$, we use the 3D nonlinear GPE
\begin{equation}
 \label{eq_TB_GPE_full3D}
	\imag \hbar \pdiff[]{}{t} \psi\qty{\mathbf{r}, t} = \sqty{-\frac{\hbar^{2}}{2m} \nabla^2 + V\qty{\mathbf{r}, t} + \frac{4\pi\hbar^{2} a_s}{m}N \abs{\psi\qty{\mathbf{r}, t}}^{2}}\psi\qty{\mathbf{r}, t}
\end{equation}
for the macroscopic wave function $\psi=\psi\qty{\mathbf{r}, t}$ which is normalized according to the condition 
\begin{equation}
 \label{eq_normalisation_condition}
    \int \dd^3 {r} \abs{\psi\qty{\mathbf{r}, t}}^{2} = 1,
\end{equation}
where $\mathbf{r} \equiv \qty{x,y,z}$ is the position vector with the Cartesian coordinates $x$, $y$ and $z$.
Here we assume that the atoms are interacting via a contact potential whose strength is determined by the $s$-wave scattering length $a_s$.

In order to realize shell-shaped quantum gases two alternatives offer themselves which have both been demonstrated experimentally: either employing rf-dressing techniques to realize a bubble potential~\cite{Zobay.2001,Garraway.2016,Carollo.2022} or taking advantage of the tuneable inter-species interaction in a dual-species BEC mixture to achieve a core-shell density distribution~\cite{Wolf.2022,Jia.2022}. In both cases the shell condensate effectively experiences a bubble potential which can be approximated by the form
\begin{equation}
 \label{eq:potential_V}
	V\qty{\mathbf{r}, t} \equiv \frac{m\omega_0^2}{2}\left[f(t)\right]^2\left(\abs{\mathbf{r}}-r_0\right)^2
\end{equation}
corresponding to a spherically symmetric harmonic potential with its minimum at the radial postition $r_0$ and the time-dependent frequency $\omega_0 f(t)$, determined by the frequency amplitude $\omega_0$ and the scaling function $f(t)$.
By changing  $r_0$ and $\omega_0$ the initial diameter and thickness of the shell can be controlled, respectively. Moreover, the time-dependent function $f(t)$ is used to dynamically adjust the width of the shell potential and is the key parameter for realizing matter-wave lensing in this article.

\subsection{Analytical model for large quantum gas shells}
\label{sec:1D_model}

The GPE is a nonlinear partial differential equation involving three spatial and one temporal variables and generally it is not easy to solve. By using the fact that the trapping potential $V(\mathbf{r},t)$, \eqref{eq:potential_V}, only depends on the radial coordinate $r\equiv\abs{\mathbf{r}}$ and time $t$, we can consider spherically symmetric solutions of \eqref{eq_TB_GPE_full3D} and therefore effectively reduce the number of relevant spatial dimensions from three to one. Still, in this case the dynamics of a BEC is in general too complicated to be described with a reliable analytical model. Hence, in order to enable an analytical description and to apply it for finding the conditions of the optimal collimation of a shell-shaped BEC, we consider in this article the case where the shell radius is much larger than the thickness of the shell. This so-called \textit{thin-shell limit} is reached when the minimum $r_0$ of the trapping potential \eqref{eq:potential_V} is much larger than the characteristic size $a_{\mathrm{HO}}\equiv\sqrt{\hbar/m\omega_0}$ of the ground state in the harmonic oscillator of frequency $\omega_0$. In this regime we can apply two approximations which simplify the underlying differential equations and enable an analytical treatment.

In \ref{Appendix:role_of_atom_atom_interaction} we have shown that for the ground state of a shell-shaped BEC the atom-atom interaction becomes negligible for large shell radius, more precisely for $r_0\gg \sqrt{Na_s a_{\mathrm{HO}}}$. In other words, this inequality means that the atom-atom interaction term in \eqref{eq_TB_GPE_full3D}, being of the order of $(4\pi\hbar^2a_{\mathrm{s}}/m)(N/V_{\mathrm{sh}})$ with the shell volume $V_{\mathrm{sh}}=4\pi r_0^2 a_{\mathrm{HO}}$, is small in comparison with the ground state energy $\hbar\omega_0$. Consequently, we can neglect the contribution from the interaction and use the 3D linear Schr\"odinger equation instead of the 3D nonlinear GPE to describe large shells. 

When additionally $r_0\gg a_{\mathrm{HO}}$, the ground state of a shell-shaped BEC can be well described by the stationary 1D Schr\"odinger equation, instead of the 3D version, as we have proven in \ref{Appendix:3D-1D}. In this case the curvature of the shell is negligible and the system can be characterized with a single Cartesian position coordinate $x\in  (-\infty,\infty)$ drastically reducing the complexity of the spatial derivatives in \eqref{eq_TB_GPE_full3D}. 

Thus,
if $r_0\gg \mathrm{max}\left\{a_{\mathrm{HO}},\sqrt{Na_sa_{\mathrm{HO}}}\right\}$, it is sufficient to apply the 1D stationary Schr\"odinger equation to obtain the ground state of the system. In fact, in many practical implementations $N a_s \gg a_{\mathrm{HO}}$ such that neglecting the interaction is the stronger requirement. As a result, the optimal regime for applying our analytical model is the \textit{non-interacting thin-shell limit} which can be accessed with rather modest experimental parameters as we have shown in \ref{Appendix:from_3D_GPE_to_1D_SE}. 

In order to model the dynamics of the shell-shaped BEC in the case of the time-dependent frequency $\omega_0 f(t)$, we use the 1D Schr\"odinger equation
\begin{equation}
 \label{eq:1d_Schroedinger_equation_result}
    i\hbar\frac{\partial}{\partial t}\varphi(x,t)=
    \left\{-\frac{\hbar^2}{2m}\frac{\partial^2}{\partial x^2}+\frac{m\omega_0^2}{2}[f(t)]^2(x-r_0)^2\right\}\varphi(x,\tau)
\end{equation}
for the wave function $\varphi(x,t)\equiv \sqrt{4\pi} x\psi(x,t)$ for $x\geq 0$, with $\psi(r,t)$ being the spherically symmetric solution of the 3D GPE (\ref{eq_TB_GPE_full3D}) and $r$ substituted by $x$. By comparing the analytical results obtained from \eqref{eq:1d_Schroedinger_equation_result} to the ones based on full 3D numerical simulations of the GPE, we show in Sec.~\ref{sec:section3} that our 1D analytical model makes correct predictions as long as the atomic density $\abs{\psi(\mathbf{r},t)}^2$ is zero around the origin, that is for $0\leq r\lesssim a_{\mathrm{HO}}$, during the whole dynamics. This requirement is equivalent to being in the non-interacting thin shell limit.

\subsection{Solutions for the dynamics of large shells}
\label{sec:1DDynamics}

Finally, we can now show that the problem of solving the partial differential \eqref{eq:1d_Schroedinger_equation_result} can be reduced to solving a linear ordinary differential equation which enables an efficient determination of the time-dependent width of the shell. In particular, according to \ref{Appendix:effective_1D_model}, the general solution of \eqref{eq:1d_Schroedinger_equation_result} can be written analytically for any initial wave function $\varphi(x,0)$. 

In our scheme, the shell-shaped BEC is initially prepared in a trapping potential $V(\mathbf{r},t)$ of the form of \eqref{eq:potential_V} with $f=1$ and $r_0\gg \mathrm{max}\left\{a_{\mathrm{HO}},\sqrt{Na_sa_{\mathrm{HO}}}\right\}$. According to \ref{Appendix:from_3D_GPE_to_1D_SE}, $\varphi(x,0)$ then coincides very well with the normalized wave function
\begin{equation}
 \label{eq:1d_Schroedinger_equation_initial wave_function}
    \varphi(x,0)=\frac{1}{\left(\sqrt{\pi} a_{\mathrm{HO}}\right)^{1/2}}\exp\left\{-\frac{(x-r_0)^2}{2a_{\mathrm{HO}}^2}\right\}
\end{equation}
of the ground state in the 1D harmonic oscillator. 

Hence, the solution of \eqref{eq:1d_Schroedinger_equation_result} corresponding to this initial state 
reads 
\begin{equation}
 \label{eq:1d_Schroedinger_equation_wave_function_result}
    \varphi(x,t)=\frac{1}{\left[\sqrt{\pi} a_{\mathrm{HO}}\lambda\right]^{1/2}}\exp\left[-\left(\frac{1}{\lambda^2}-i\frac{\dot{\lambda}}{\omega_0\lambda}\right)\frac{(x-r_0)^2}{2a_{\mathrm{HO}}^2}-\frac{i}{2}\Phi\right] ,
\end{equation}
where 
\begin{equation}
    \Phi(t)=\omega_0\int_{0}^{t}\frac{dt'}{[\lambda(t')]^2}
\end{equation}
is a time-dependent phase factor and $\lambda=\lambda(t)$ with $\dot{\lambda}\equiv d\lambda/dt$. 

The time-dependent function $\lambda(t)$ has to fulfill the corresponding Ermakov equation
\begin{equation}
 \label{eq:Ermakov_equation_definition}
    \frac{d^2}{dt^2}\lambda+[\omega_0 f(t)]^2\lambda=\frac{\omega_0^2}{\lambda^3}
\end{equation}
with the initial condition $\lambda(0)=1$ and $\dot{\lambda}(0)=0$. 

Although the Ermakov equation (\ref{eq:Ermakov_equation_definition}) is a nonlinear ordinary differential equation of second order, its solution can be represented in terms of solutions of the corresponding linear differential equation. Indeed, as shown also in \ref{Appendix:effective_1D_model}, the solution of \eqref{eq:Ermakov_equation_definition} is given by
\begin{equation}
 \label{eq:Ermakov_equation_solution_result}
    \lambda(t)=\sqrt{\left[\Lambda_1(t)\right]^2+\left[\Lambda_2(t)\right]^2}, 
\end{equation}
where $\Lambda_1(t)$ and $\Lambda_2(t)$ are two linearly independent solutions of the linear differential equation 
\begin{equation}
 \label{eq:equation_for_lambda_12}
    \frac{d^2}{dt^2}\Lambda+[\omega_0 f(t)]^2\Lambda=0
\end{equation}
with the initial conditions $\Lambda_1(0)=1$ and $\dot{\Lambda}_1(0)=0$, and $\Lambda_2(0)=0$ and $\dot{\Lambda}_2(0)=\omega_0$. 

From a physics view-point, the function $\lambda(t)$ defines the dependence of the variance 
\begin{equation}
 \label{eq:variance_result}
    \sigma^2 (t)\equiv \langle x^2\rangle - \langle x\rangle ^2=\int_{-\infty}^{+\infty} x^2|\varphi(x,t)|^2 dx - r_0^2=\frac{1}{2}[a_{\mathrm{HO}}\lambda(t)]^2
\end{equation}
of the position distribution $\abs{\varphi(x,t)}^2$, with $\varphi(x,t)$ given by \eqref{eq:1d_Schroedinger_equation_wave_function_result}, on time. Here we have used the fact that the mean value of the position 
\begin{equation}
 \label{eq:mean_value_result}
    \langle x\rangle (t)\equiv\int_{-\infty}^{+\infty} x|\varphi(x,t)|^2 dx =r_0
\end{equation}
is time-independent and coincides with the minimum of the trapping potential. 

Thus, the dynamics of a shell-shaped BEC trapped in the 3D harmonic potential \eqref{eq:potential_V} with large radius $r_0$, and in particular the time evolution of the shell width $\sigma(t)$, is solely determined by the solutions $\Lambda_1(t)$ and $\Lambda_2(t)$ of the linear differential equation (\ref{eq:equation_for_lambda_12}) for a given frequency profile $f(t)$. In the following we apply this model to efficiently describe matter-wave lensing of shell-shaped BECs and to obtain optimal parameters for future experimental implementations.    

\section{Matter-wave lensing of shell-shaped Bose-Einstein condensates}
\label{sec:section3}

In this section we analyze matter-wave lensing of shell-shaped BECs by applying the conventional delta-kick collimation (DKC) technique. Based on our 1D analytical model, we then derive the optimal parameters which allow us to keep both the shell radius and width almost constant for the longest possible free expansion times. Finally, we consider a more generalized matter-wave lensing scheme and discuss its applicability.

\subsection{Delta-kick collimation with shell potentials}
\label{sec:section3p1}

We start our analysis of shell-shaped BEC collimation from reminding the basic principles of the widely utilized scheme of DKC~\cite{Chu1986,Ammann1997,Kovachy2015,Pandey_2021,Deppner_2021}. Indeed, this scheme consists of three steps: (i) a BEC prepared in a 3D harmonic potential is released from the trap at $t=0$ and experiences a free expansion during the delay time $t_d$, (ii) the initial harmonic potential is turned on again for a short kick time $t_k$, and (iii) is switched off completely afterwards so that the BEC evolves freely. During the first step the BEC expands freely and increases its spatial size, accompanied by a conversion of interaction energy into kinetic energy. In the second step, the wave function of the BEC picks up a position-dependent phase proportional to the harmonic potential, similar to the well-known thin lens in optics. In this way, fast particles get a larger kick towards the center of the condensate compared to slower ones, resulting in a reduction of the width of the momentum distribution and consequently slowing down the expansion of the BEC.

Here we apply the DKC scheme to collimate a shell-shaped BEC by considering the following time profile
\begin{equation}
 \label{eq:profile_DKC}
    f_{\mathrm{DKC}}(t)=
    \begin{cases}
    0,\;\;\; 0<t\leq t_d\\
    1, \;\;\; t_d<t\leq t_d+t_k\\
    0, \;\;\; t_d+t_k < t
    \end{cases}
\end{equation}
for the trapping frequency of the bubble potential \eqref{eq:potential_V}. Indeed, the lensing scheme is fully determined by the delay time $t_d$ and the kick time $t_k$ while the strength of the kicking potential is chosen equally to the initial potential without loss of generality (see Sec.~\ref{sec:section3p3}). 

\begin{figure}
	\centering
	\includegraphics{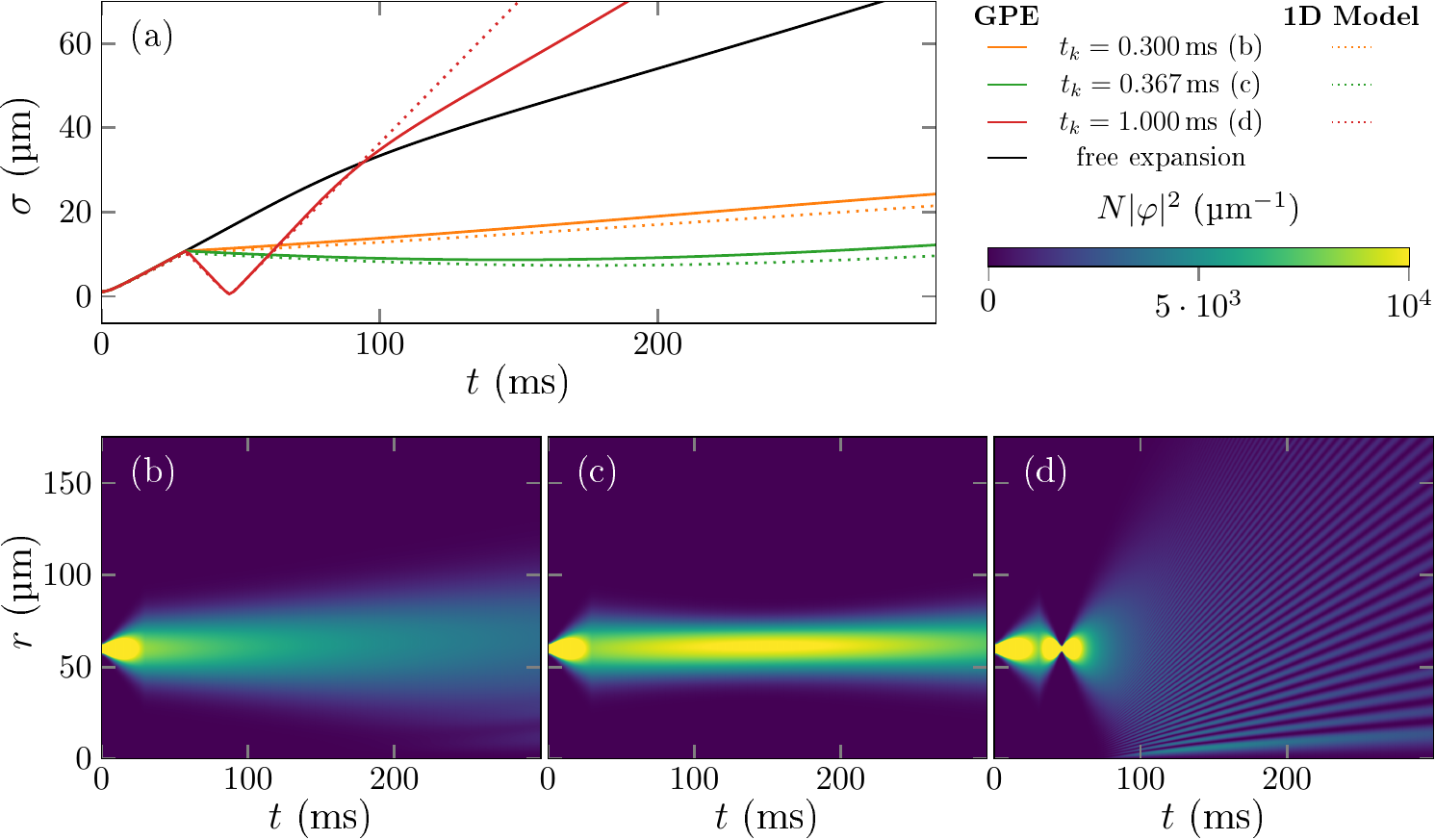}
	\caption{(a) Dependence of the shell width $\sigma(t)$ on the expansion time $t$ obtained by numerical simulations of the 3D GPE, \eqref{eq_TB_GPE_full3D}, (solid lines) compared to the analytical 1D model, \eqref{eq:variance_result}, (dotted lines) for the time profile $f(t)=f_{\mathrm{DKC}}(t)$, \eqref{eq:profile_DKC}, with delay time $t_d=30\,\mathrm{ms}$ and different values of the kick time $t_k$ including the purely free expansion (black). 
	(b), (c) and (d) Time evolution of the radial distribution $|\varphi(r,t)|^2\equiv 4\pi r^2|\psi(r,t)|^2$ for each of the nonzero values of $t_k$, respectively.\label{fig:DKC_VARR_densitys}}
\end{figure}

In \figref{fig:DKC_VARR_densitys} we present the results of the delta-kick scheme, \eqref{eq:profile_DKC}, with $t_d=30\,\mathrm{ms}$ and for different values of $t_k$, applied to the typical case of a $^{87}\mathrm{Rb}$ BEC with $N=10^5$ atoms, $s$-wave scattering length $a_s=5.29\, \mathrm{nm}$, and trapping potential $V(\mathbf{r},t)$ given by \eqref{eq:potential_V} with parameters $f(0)=1$, $r_0=60$\,\textmu m and $\omega_0=2\pi\cdot 50\,\mathrm{Hz}$.
The solid lines in \figref{fig:DKC_VARR_densitys} (a) represent the time evolution of the standard deviation of the shell $\sigma(t)$ obtained by solving the 3D GPE (\ref{eq_TB_GPE_full3D}) numerically and correspond to the width of the radial density distributions plotted in \figref{fig:DKC_VARR_densitys} (b), (c), and (d).

Obviously, for $t_k=0$, the shell only spreads up freely, as depicted by the solid black line in \figref{fig:DKC_VARR_densitys} (a). 
For $t_k=0.3\,\mathrm{ms}$ (orange) we already observe a considerable slow down of the expansion, but in this case the shell width is still growing monotonically corresponding to a too short, undershooting lens. 
By further increasing the kick time the lens leads to an overshooting such that the width of the shell decreases directly after the delta-kick and then starts to grow once it reached its minimum extension. This behavior is exemplarily displayed for $t_k=1.0\,\mathrm{ms}$ (red). 
The compromise between both regimes is given by the optimal kick time $t_k=0.367\,\mathrm{ms}$ (green) which only slightly overshoots and therefore keeps the shell width almost constant for several $100\,\mathrm{ms}$ of free expansion time. 
As we will show in more detail in the next section, there is indeed a certain value of $t_k$ for a given delay time $t_d$ leading to an optimal DKC performance such that the shell keeps its width for a long time without any external potential.

As the next step of our analysis, we quantitatively compare the results of the exact 3D numerical simulations with the prediction of the 1D model presented in Sec.~\ref{sec:1DDynamics}. In \ref{Appendix:effective_1D_model} we have solved \eqref{eq:equation_for_lambda_12} with $f(t)$ given by \eqref{eq:profile_DKC} and derived the analytical formulas for the functions $\Lambda_1 (t)$ and $\Lambda_2 (t)$. With the help of Eqs. (\ref{eq:Ermakov_equation_solution_result}) and (\ref{eq:variance_result}) we then obtain the shell width $\sigma(t)$ and present its time dependence in \figref{fig:DKC_VARR_densitys} (a) by dotted lines for the same values of $t_k$. As a result, the solid and dotted lines are very close to each other as long as the width stays below $40$\,\textmu m, meaning that the 1D model describes the shell dynamics quite well in this regime. 
This behavior is in good agreement with the considerations of Sec.~\ref{sec:1D_model} requiring the shell width to stay below the initial shell radius $r_0 = 60\,\text{\textmu m}$. Obviously this condition gets violated for the strongly overshooting lens (red curve) leading to deviations between the 1D model and the full 3D numerical simulation. 

Moreover, slight deviations between the two approaches are due to the atom-atom interaction which is not taken into account by the 1D model. Indeed, neglecting the interaction requires $r_0/a_{\mathrm{HO}}\gg \sqrt{Na_s/a_{\mathrm{HO}}}$. For our parameters we have $r_0/a_{\mathrm{HO}}\approx 39$ and $\sqrt{Na_s/a_{\mathrm{HO}}}\approx 19$, such that the interaction still plays a minor, but relevant role. 

Finally, we can conclude that the 1D model predicts correct results as long as $\sigma (t)\ll r_0$ and the atom-atom interaction is not too strong.

\subsection{Conditions for optimal lensing parameters}
\label{sec:optimal}

The results presented in \figref{fig:DKC_VARR_densitys} show that the DKC method works very well for keeping the width $\sigma(t)$ of a shell-shaped BEC at an almost constant value during the free evolution time $t$ when slightly overshooting the lens. Now we are interested in obtaining specific conditions for the timings $t_d$ and $t_k$ to extend this time as long as possible in future experiments.

A suitable measure to characterize the performance of the lens is then given by the time interval $\Delta t$, which corresponds to the free expansion time after the lens that is required until the shell BEC reaches again the width it initially had at the time of the lens $t_d+t_k$. Formally, $\Delta t$ can thus be defined by the relation
\begin{equation}
 \label{eq:definition_of_deltat}
    \sigma(t_d+t_k+\Delta t) = \sigma(t_d+t_k).
\end{equation}
A comparison with \figref{fig:DKC_VARR_densitys} reveals that $\Delta t$ is rather short for a heavily overshooting lens (red curve) and maximal for slightly overshooting (green curve).

By inserting the analytical formulas for the functions $\Lambda_1(t)$ and $\Lambda_2(t)$, derived in \ref{Appendix:effective_1D_model}, into \eqref{eq:Ermakov_equation_solution_result}, we obtain from Eqs. (\ref{eq:variance_result}) and (\ref{eq:definition_of_deltat})
\begin{equation}
 \label{eq:result_for_deltat}
    \Delta t (t_d,t_k)=t_d\,\frac{\omega_0 t_d\sin(2\omega_0 t_k)-2\cos(2\omega_0 t_k)}{\left[\sin(\omega_0 t_k)\right]^2+\left[\omega_0 t_d\sin(\omega_0 t_k)-\cos(\omega_0 t_k)\right]^2}.
\end{equation}
We note that the right-hand side of \eqref{eq:result_for_deltat} might be negative for certain values of $t_d$ and $t_k$ and we therefore set $\Delta t$ equal to zero in these nonphysical situations. 

For a given $t_d$, $\Delta t$ reaches its maximum value when $t_k$ is given by
\begin{equation}
 \label{eq:result_for_deltat_maximum}
    t_k^{(n)}=\frac{\pi n}{\omega_0}+\frac{1}{2\omega_0}\left[\frac{\pi}{2}+\arctan\left(\frac{2}{\omega_0t_d}\right)-\arctan\left(\frac{\omega_0t_d}{2}\sqrt{\omega_0^2 t_d^2+4}\right)\right] ,
\end{equation}
with $n=0,1,2...$. The function $t_k^{(n)}(t_d)$ is a multi-valued one, resulting from the fact that $\Delta t$, \eqref{eq:result_for_deltat}, is a periodic function of $t_k$ with period $\pi/\omega_0$. Here we consider only the case $n=0$ yielding the maximum value 
\begin{equation}
 \label{eq:result_for_deltat_maximum_function_of_td}
    \Delta t_{\mathrm{max}}^{\mathrm{DKC}}\equiv \Delta t (t_d,t_k^{(n)})=t_d\sqrt{\omega_0^2 t_d^2+4}
\end{equation}
of the time interval $\Delta t$, \eqref{eq:result_for_deltat}, which is an increasing function of $t_d$. 

As an example, for the delay time $t_d=30\,\mathrm{ms}$ used in \figref{fig:DKC_VARR_densitys}, we obtain from Eqs. (\ref{eq:result_for_deltat_maximum}) and (\ref{eq:result_for_deltat_maximum_function_of_td}) the optimal kick time $t_k^{(0)}=0.367\,\mathrm{ms}$ corresponding to the time interval $\Delta t_{\mathrm{max}}=289\,\mathrm{ms}$.

\begin{figure}[t]
	\centering
	\includegraphics{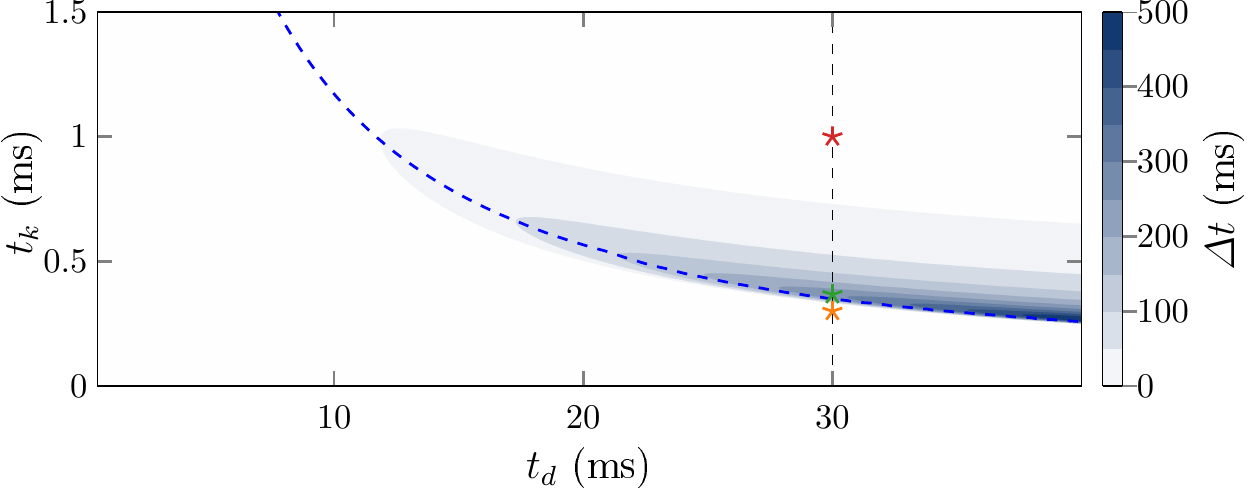}
	\label{fig:DKC_parameterscan}
	\caption{Contour plot of the time interval $\Delta t$, defined by \eqref{eq:definition_of_deltat} and given by \eqref{eq:result_for_deltat}, for different holding $t_d$ and kick $t_k$ times of the time profile $f_{\mathrm{DKC}}(t)$, \eqref{eq:profile_DKC}. The three stars on the dashed black line mark the values of $t_d$ and $t_k$ corresponding to the subfigures (b), (c) and (d) of \figref{fig:DKC_VARR_densitys}. The blue dashed line identifies the optimal values of $t_k$, \eqref{eq:result_for_deltat_maximum} with $n=0$, as a function of $t_d$, resulting in the maximum value of $\Delta t$, \eqref{eq:result_for_deltat_maximum_function_of_td}. 
    \label{fig:DKC_parascan}
	}
\end{figure}

In \figref{fig:DKC_parascan} the dependence of the time interval $\Delta t$, \eqref{eq:result_for_deltat}, on the delay time $t_d$ and kick time $t_k$ is visualized in a contour plot. In addition, the optimal kick time $t_k^{(0)}$, \eqref{eq:result_for_deltat_maximum}, is also displayed by the dashed blue line.
As a result, the collimation of a shell is better for larger $t_d$ and smaller $t_k$ times.
However, in order to realize a thin and almost non-spreading shell for a long time, the free expansion time $t_d$ should not be too large. In particular, for longer $t_d$, $\omega_0 t_d\gtrsim 1$, the thickness of the shell $\sigma$ increases linearly with $t_d$, namely $\sigma(t_d)\approx a_{\mathrm{HO}}\,\omega_0 t_d/\sqrt{2}$, as shown in \ref{Appendix:effective_1D_model}. Hence, in order to ensure $\sigma(t_d)\ll r_0$ with $r_0\gg a_{\mathrm{HO}}$, the delay time $t_d$ should obey the inequality $1\lesssim \omega_0 t_d \ll r_0/a_{\mathrm{HO}}$.  

Consequently, a thin and slowly spreading shell can be realized in an experiment by utilizing the standard DKC scheme, corresponding to the time-profile $f_{\mathrm{DKC}}(t)$, \eqref{eq:profile_DKC}, of the trapping frequency, provided the delay time $t_d$ and kick time $t_k$ obey the optimal condition \eqref{eq:result_for_deltat_maximum} with $n=0$, displayed by blue dashed line in \figref{fig:DKC_parascan}. In this case the shell width stays below a certain threshold during the time interval $\Delta t_{\mathrm{max}}^{\mathrm{DKC}}$, \eqref{eq:result_for_deltat_maximum_function_of_td}, enabling the study of freely floating shells for extended times.

\subsection{Generalized matter-wave collimation}
\label{sec:section3p3}

As we will discuss in more detail in section \ref{sec:section4} it is not possible in all physical setups to switch off the trapping potential completely before performing the delta-kick pulse. Therefore, we now present a generalization of the conventional DKC technique that consists of the following three steps: (i) for $0<t<t_d$, the frequency $\omega_0 f(t)$ of the trap potential $V$, \eqref{eq:potential_V}, is not set to zero but rather to a finite constant value $\omega_0 f_1$ with $f_1>0$, and (ii) for $t_d<t<t_d+t_k$, the trapping frequency is again changed to another value $\omega_0 f_2$ with $f_2>0$, followed by (iii) turning off the trapping potential completely, for $t>t_d+t_k$. Hence, our general scheme is described by the time profile
\begin{equation}
 \label{eq:general_profile_main}
    f_{\mathrm{gDKC}}(t)=
    \begin{cases}
    f_1,\;\;\; 0<t\leq t_d\\
    f_2, \;\;\; t_d<t\leq t_d+t_k\\
    0, \;\;\; t_d+t_k < t
    \end{cases}
\end{equation}
of the trapping frequency. When choosing $f_1 = 0$ and $f_2 = 1$, sequence \eqref{eq:general_profile_main} coincides with \eqref{eq:profile_DKC} and we obtain the previously discussed DKC scheme. 

By using these new parameters $f_1$ and $f_2$, we analyze how to best slow down the spreading of the shell-shaped BEC in comparison with the conventional DKC technique considered in section~\ref{sec:optimal} when switching off the potential completely is not appropriate for the system under study. In order to find the optimal values for $f_1$ and $f_2$, we have again used the time interval $\Delta t$, defined by \eqref{eq:definition_of_deltat}, and maximized it with respect to $t_k$ and $t_d$ for given values of $f_1$ and $f_2$. As a result, $\Delta t$ achieves its maximal value
\begin{equation}
 \label{eq:result_for_deltat_maximum_function_of_f1f2}
    \Delta t_{\mathrm{max}}^{\mathrm{gDKC}}(f_1,f_2)=\frac{|f_1^4-f_2^2|}{f_1^2 f_2^2 \omega_0}
\end{equation}
at the optimal values 
\begin{equation}
 \label{eq:result_for_deltat_maximum_f1f2_optimaltd}
    t_d^{(n_1)}=\frac{\pi}{2\omega_0 f_1}(2n_1+1)
\end{equation}
and
\begin{equation}
 \label{eq:result_for_deltat_maximum_f1f2optimal_tk}
    t_k^{(n_2)}=\frac{1}{2\omega_0 f_2}\left[2\pi n_2+\pi\Theta(f_1^2-f_2)+\arccos\left(\frac{|f_1^4-f_2^2|}{f_1^4+f_2^2}\right)\right]
\end{equation}
of the time $t_k$ and $t_d$ for given $f_1$, and $f_2$, where $n_{1,2}=0,1,2...$ and $\Theta(z)$ is the Heaviside step function, $\Theta(x> 0)=1$ and $\Theta(x<0)=0$. 

Now we can compare the efficiency of the generalized and standard DKC schemes to keep the shell width almost constant for longer free expansion time. Obviously, for realizing long-lived shells, one should avoid the case $f_1^2=f_2$ which leads to $\Delta t_{\mathrm{max}}^{\mathrm{gDKC}} = 0$ according to \eqref{eq:result_for_deltat_maximum_function_of_f1f2}. In general there are two main scenarios depending on the relation between $f_1$ and $f_2$:

(i) For $f_1^2\ll f_2$, we obtain $\Delta t_{\mathrm{max}}^{\mathrm{gDKC}}\approx 1/(\omega_0 f_1^2)$ and can distinguish two cases. The first case, $f_1^2\ll f_2$ and $f_1\ll 1$, resembles the standard DKC approach and the optimal values of $t_d$, \eqref{eq:result_for_deltat_maximum_f1f2_optimaltd}, and $t_k$, \eqref{eq:result_for_deltat_maximum_f1f2optimal_tk}, are given by $t_d^{(0)}\rightarrow \infty$ and $t_k^{(0)}\rightarrow 0$. They agree with the results derived in Sec.~\ref{sec:optimal}. and presented in \figref{fig:DKC_parascan}. Moreover, the maximal time interval $\Delta t_{\mathrm{max}}^{\mathrm{gDKC}}\approx 1/(\omega_0 f_1^2)=(2/\pi)^2\omega_0 (t_d^{(0)})^2$ is approximately the same as the one of the standard DKC scheme, $\Delta t_{\mathrm{max}}^{\mathrm{DKC}}=\omega_0 t_d^2$, \eqref{eq:result_for_deltat_maximum_function_of_td} for $\omega_0 t_d\gg 1$. In the second case, $f_1^2\ll f_2$ and $f_1\geq 1$, the generalized DKC scheme becomes less efficient compared to the standard one because the denominator of \eqref{eq:result_for_deltat_maximum_function_of_f1f2} grows when increasing $f_1$ beyond one. 

(ii) For $f_1^2\gg f_2$, we obtain $\Delta t_{\mathrm{max}}^{\mathrm{gDKC}}\approx (1/\omega_0)(f_1/f_2)^2$. As a result, it is better to have $f_1\gg f_2$, rather than $f_1\ll f_2$. The former case can be realized for both small, $f_2<1$, and large, $f_2\geq 1$, values of $f_2$. 

To summarize, the generalized DKC scheme does not necessarily lead to a performance gain for matter-wave lensing. However, it enables lensing for a broader set of physical systems, in particular for the case of rf dressing discussed in the next section.

\section{Application to rf-dressed potentials}
\label{sec:section4}

In this section we consider the case of rf-dressing for generating shell-shaped BECs and propose a realistic matter-wave lensing scheme enabling long-lived shells that takes into account all relevant experimental aspects. 

\subsection{Radio frequency dressed potentials}

Nowadays, a promising approach to create shell-shaped BECs is to employ the rf potential \cite{Zobay.2001,Zobay.2004,Garraway.2016,Perrin.2017} given by
\begin{equation}
    \label{eq:rf-potential}
    V_{\mathrm{rf}}(r) = M_F\frac{g_F}{|g_F|} \sqrt{\left(\frac{m \omega_{M}^2}{2F} r^2 - \hbar \Delta \right)^2 + \left(\hbar \Omega_0\right)^2},
\end{equation}
where $M_F$ is the momentum projection of the total momentum $F$ of a dressed state in the hyperfine manifold with
corresponding Land\'e factor $g_F$. The trap frequency $\omega_{M}$ of the static magnetic trap is chosen such that the potential of the highest trapped bare state is given by
$V_{\mathrm{st}}(r) = m\omega_M^2/2$. Moreover, $\Delta$ is the detuning of the rf field with respect to the transition between neighboring bare states at the center of the trap, and $\Omega_0$ is the Rabi frequency proportional to the magnitude of the rf field. 

The potential, \eqref{eq:rf-potential}, can be expanded around its minimum
\begin{equation}
\label{eq:rf_r0}
  r_0 = \left(2F\frac{\hbar\Delta}{m \omega_{M}^2}\right)^{1/2}
\end{equation}
which yields
\begin{equation}
    \label{eq:rfpot_taylor}
   V_{\mathrm{rf}}(r) \approx M_F\frac{g_F}{|g_F|} \hbar \Omega_0 + \frac{m\omega_0^2}{2} (r-r_0)^2
 \end{equation}
with the local trap frequency
\begin{equation}
\label{eq:rf_omega0}
  \omega_0 = \omega_{M}\left(\frac{2M_F g_F}{F|g_F|} \frac{\Delta}{\Omega_0}\right)^{1/2}.
\end{equation}

Thus, the rf potential $V_{\mathrm{rf}}(r)$ can be approximated by a spherically symmetric harmonic oscillator shifted along the coordinate $r$ by $r_0$, that is the rf potential has the form of \eqref{eq:potential_V}. Therefore, we might utilize the results based on the standard DKC scheme to an rf-dressed potential. However, the DKC technique would not work, since a sudden switch off and on of the potential result in a mixture of different $M_F$ states leading to atom losses in the shell. Furthermore, changing the trapping frequency $\omega_0$ too fast in time is also detrimental since the $M_F$ states need to follow the local magnetic field to feel the intended potential.
To resolve these issues of employing an rf-dressed potential we present in the next section an adapted DKC approach that involves a continuous change of the trap frequency rather than sudden jumps.

\subsection{Scheme and results for rf-dressed bubbles}
\label{sec:section4.2_RF_results}

To perform matter-wave lensing of a shell-shaped BEC initially created by rf-dressing with the potential (\ref{eq:rfpot_taylor}) which has the form of \eqref{eq:potential_V}, we model the time profile $f(t)$ with the following relation 
\begin{equation}
 \label{eq: profile RF}
    f_{\mathrm{RF}}(t)=
    \begin{cases}
    1 - (1-f_1)(t/t_r),\;\;\; 0<t\leq t_r\\
    f_1, \;\;\; t_r<t\leq t_r+t_d\\
    0, \;\;\; t_r+t_d < t
    \end{cases}
\end{equation}
which is an adaption of the generalized DKC scheme discussed in section \ref{sec:section3p3}.
Here the sequence starts with a linear ramp during the time $0<t<t_r$ to reduce the trap frequency from $\omega_0$ to $f_1 \omega_0$, which is then followed by keeping the trap frequency constant for the time $t_d$ until the trap is finally turned off at $t=t_r+t_d$. 
The linear ramp ensures that the trap parameters are changed slow enough to not mix different $M_F$ states. While keeping the trap constant afterwards with a reduced trap frequency the BEC will expand in that shallower trap and excite breathing oscillations. When the trap is switched off at the time the cloud has reached its maximum width, and therefore maximum potential energy, a huge amount of energy can be removed from the system leading to a slow free expansion of the cloud. 
This concept resembles the original DKC proposal by Chu et al.~\cite{Chu1986}.

\begin{figure}[t]
	\centering
	\includegraphics{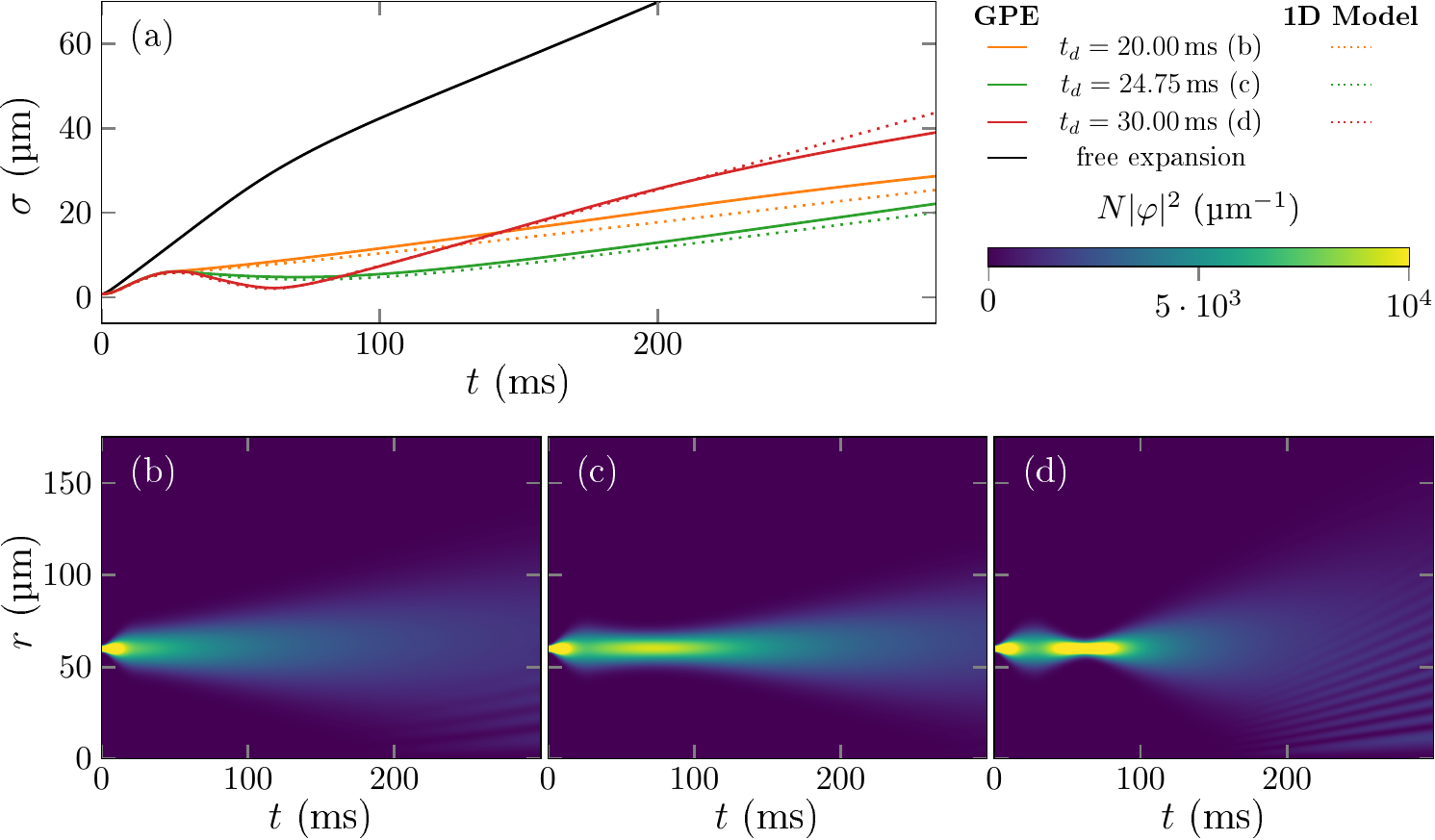}
	\caption{(a) Dependence of the shell width $\sigma(t)$, \eqref{eq:variance_result}, on the expansion time $t$ obtained by solving numerically the 3D GPE (solid lines) compared to the results of the analytical 1D model (dotted line) with the time profile $f(t)=f_{\mathrm{RF}}(t)$, \eqref{eq: profile RF}, for $f_1=0.1$, $t_r=5\,\mathrm{ms}$, and different values of $t_d$.
	For comparison the black solid line results from the numerical solution of the 3D GPE with no ramp and $t_d=0$. (b), (c) and (d) Time evolution of the radial distribution $|\varphi(r,t)|^2\equiv 4\pi r^2|\psi(r,t)|^2$ for each of the nonzero values of $t_d$, respectively.
	\label{fig:RF_VARR_densitys}
	}
\end{figure}

We have studied the performance of the lensing scheme with $f(t)=f_{\mathrm{RF}}(t)$, \eqref{eq: profile RF}, by carrying out numerical simulations of the 3D GPE (\ref{eq_TB_GPE_full3D}) with ramping time $t_r = 5\,\mathrm{ms}$ and different values of $t_d$.
The results of the time evolution of the shell width $\sigma(t)$ are displayed by the corresponding solid lines in \figref{fig:RF_VARR_densitys} (a) and have been obtained for the parameters $r_0 = 61$\,\textmu m, $\omega_0 = 2\pi \cdot 100\,\mathrm{Hz}$ and $f_1=0.1$. The dashed lines are results of the 1D model, presented in Sec. 2, with the analytical formulas for $\Lambda_{1,2}(t)$ derived in \ref{Appendix:effective_1D_model}.  
In addition, for each of the nonzero values of $t_d$, we show in \figref{fig:RF_VARR_densitys} (b), (c), and (d) the time evolution of the radial distribution $|\varphi(r,t)|^2\equiv 4\pi r^2|\psi(r,t)|^2$. 

By increasing $t_d$, we observe a transition from an undershooting lens, $t_d = 20\,\mathrm{ms}$, to an optimal collimation, $t_d=24.75\,\mathrm{ms}$, and finally to an overshooting of the lens, $t_d=30\,\mathrm{ms}$, for the adapted DKC scheme indicated by the orange, green, and red lines, accordingly. In the optimal case of $t_d=24.75\,\mathrm{ms}$, the shell keeps its shape for around $100\,\mathrm{ms}$ which is sufficient time to manipulate the shell with external fields and observe its response during free expansion. 

\begin{figure}[t]
	\centering
	\includegraphics{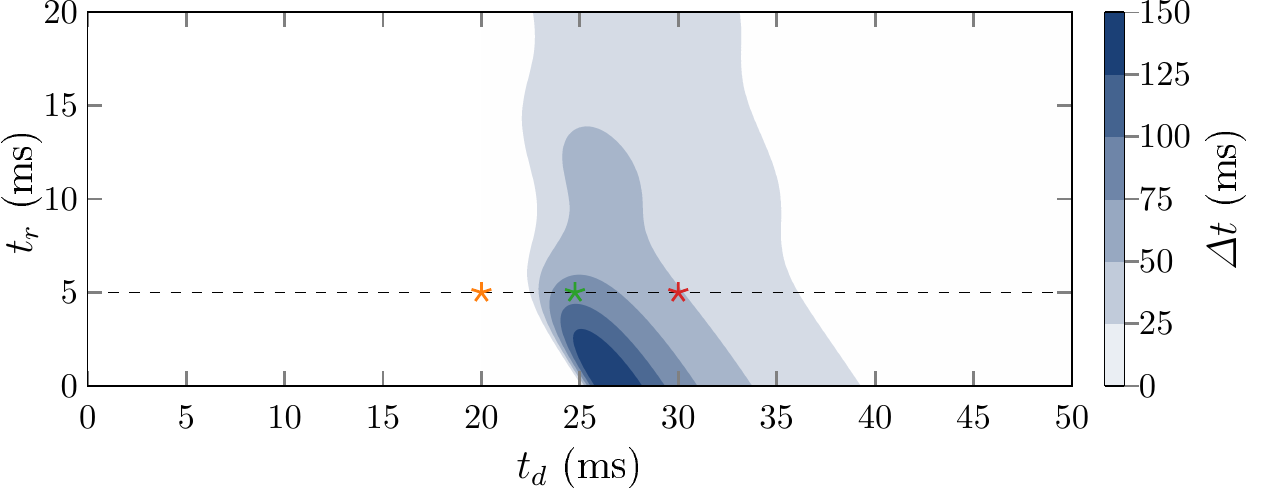}
	\caption{Contour plot of the time interval $\Delta t$, defined by \eqref{eq:definition_of_deltat}, for different ramping $t_r$ and delay $t_d$ times of the time profile $f_{\mathrm{RF}}(t)$, \eqref{eq: profile RF}. Dark colors indicate parameters leading to long free evolution times of the shell BEC. The three stars on the dashed black line mark the values of $t_r$ and $t_d$ corresponding to the subfigures (b), (c) and (d) of \figref{fig:RF_VARR_densitys}.
	\label{fig:RF_parameterscan}}
\end{figure}

To obtain the optimal values of the ramping $t_r$ and delay $t_d$ times as functions of the other system parameters, we again maximize the time $\Delta t$, defined by \eqref{eq:definition_of_deltat}. By employing the analytical solutions $\Lambda_{1,2}(t)$ derived in \ref{Appendix:effective_1D_model} for the time profile $f_{\mathrm{RF}}(t)$, \eqref{eq: profile RF}, we can express $\Delta t$ as a function of $t_r$ and $t_d$ as displayed in \figref{fig:RF_parameterscan}. Since $\Delta t$ is a periodic function of $t_d$ with period $\pi/(f_1\omega_0)$, the interval for $t_d$ in \figref{fig:RF_parameterscan} is restricted to one period. The three stars mark the values of $t_r$ and $t_d$ corresponding to subfigures (b), (c), and (d) of \figref{fig:RF_VARR_densitys}, accordingly. As a result the largest values of $\Delta t$ occur for small ramping times and $t_d\approx \pi/(2f_1\omega_0)= 25\,\mathrm{ms}$. However, as discussed above, $t_r$ should not be taken too small for the rf potential to avoid mixing the $M_F$ states during the sequence.

\subsection{Experimental feasibility}

Now we discuss more precisely the experimental details of realizing the proposed adapted DKC scheme. The shell-shaped BEC of $^{87}$Rb atoms in the state $\left|F=2,M_F=2\right\rangle$ with $|g_F|=g_F$ is prepared as the ground state of the rf potential with the trapping frequency $\omega_0 = 2\pi \cdot 100\,\mathrm{Hz}$ and the minimum $r_0=61$\,\textmu m.
According to Eqs. (\ref{eq:rf_r0}) and (\ref{eq:rf_omega0}), this corresponds to $\Delta = 2\pi \cdot 20\,\mathrm{kHz}$, $\omega_{M}= 2\pi\cdot 50\,\mathrm{Hz}$, and $\Omega_0 = 2\pi\cdot 10\,\mathrm{kHz}$. In general, there are several ways of realizing the linear ramp of the trapping frequency $f_{\mathrm{RF}}(t)$, \eqref{eq: profile RF}, while keeping $r_0$ constant at the same time. 

One option is to increase only the Rabi frequency $\Omega_0$ over time. In this way, as follows from \eqref{eq:rf_omega0}, the Rabi frequency at the end of the ramp is given by $\Omega_0(t_r)=\Omega_0(0)/f_1^2$. For $f_1=0.1$ and $\Omega_0 (0)= 2\pi\cdot 10\,\mathrm{kHz}$, we obtain $\Omega_0 (t_r)= 2\pi\cdot 1\,\mathrm{MHz}$ which seems to be too large even for state-of-the-art experiments with ultra-cold quantum gases. Thus, this approach might not be feasible.

A second option is to keep $\Omega_0$ constant, but change both $\Delta$ and $\omega_M$ while keeping the ratio $\Delta/\omega_M^2$ constant during the ramp.
Consequently, the static magnetic trap frequency $\omega_M$ and the rf detuning $\Delta$ need to follow the ramps
\begin{equation}
 \label{eq:profile_RF_omega_M}
    \omega_M^2(t)=\omega_M^2(0)\left[1-(1-f_1)\frac{t}{t_r}\right]
\end{equation}
and 
\begin{equation}
 \label{eq:profile_RF_Delta}
    \Delta(t)=\Delta(0)\left[1-(1-f_1)\frac{t}{t_r}\right].
\end{equation}

With $\omega_M(0)=2\pi\cdot 50\,\mathrm{Hz}$ and $\Delta(0)=2\pi\cdot 20\,\mathrm{kHz}$, we obtain $\omega_M(t_r)=\sqrt{f_1}\omega_M(0)\approx 2\pi\cdot 15.8\,\mathrm{Hz}$ and $\Delta(t_r)=f_1\Delta(0)=2\pi\cdot 2\,\mathrm{kHz}$ at the end of the ramp. These values correspond to typical experimental parameters and therefore a successful implementation of this scheme seems feasible.

\section{Conclusion}
\label{sec:section5}

In this article we have studied different matter-wave lensing schemes to conserve the shape of a shell-shaped BEC as long as possible during its free expansion. These techniques enable shell-shaped ultracold atomic gases to be a reliable platform for exploring many-body quantum phenomena on curved manifolds. 

By considering the case where the shell radius is large enough in comparison with the shell width and for not too strong interaction strengths, we have employed an effective 1D model, using the 1D Schr\"odinger equation instead of the 3D GPE, to describe both the ground state and dynamics of a spherically symmetric shell-shaped BEC. This model allows us to derive analytic results for an arbitrary time profile of the trapping frequency. In particular, for the conventional DKC scheme, corresponding to a step-wise change of the trapping frequency, we have obtained the optimal delay and kick times as functions of the experimental parameters for which the shell width keeps almost constant during its free dynamics. For typical parameters of state-of-the-art experiments with atomic BECs, expansion times of several hundred milliseconds can be reached.

In addition, we have proposed an adapted DKC scheme, where the trapping frequency is not changed suddenly but continuously. This technique is particularly important for shell-shaped BECs prepared with and controlled by an rf-dressed potential, where a step-wise change of the frequency would result in undesirable mixing of the different magnetic hyperfine sublevels. By obtaining optimal conditions for such a collimation scheme, we have shown that the shell structure can indeed be conserved for about hundred milliseconds.

Thus, we are confident that the results presented in our paper will boost the accessibility of freely floating shell-shaped BECs independent of the employed experimental platform and drive future advancement in quantum many-body physics. 
Furthermore, the discussed techniques can also be extended to rings and other non-trivially shaped BECs for extended application.

\ack
We thank Naceur Gaaloul and Nathan Lundblad for fruitful discussions and helpful feedback about the manuscript. This project is supported by the German Space Agency (DLR) with funds provided by the Federal Ministry for Economic Affairs and Climate Action (BMWK) due to an enactment of the German Bundestag under Grants Nos. 50WM1862 and 50WM2245B (CAL).

\appendix

\section{From 3D GPE to 1D Schrödinger equation}
\label{Appendix:from_3D_GPE_to_1D_SE}

In this Appendix we show that for a spherically symmetric harmonic oscillator with a large radial shift of its minimum the ground state of a resulting shell-shaped BEC can be well described by a 1D Schrödinger equation instead of a full 3D GPE.

\subsection{The role of atom-atom interaction in shell-shaped BECs}
\label{Appendix:role_of_atom_atom_interaction}

By introducing the dimensionless time, $\tau\equiv \omega_0 t$, and distance, $\rho\equiv r/a_{\mathrm{HO}}$, with the characteristic length $a_{\mathrm{HO}}\equiv \sqrt{\hbar/m\omega_0}$, as well as using the fact that the potential $V(\mathbf{r},t)$, \eqref{eq:potential_V}, depends only on $r$, we arrive at the dimensionless 3D GPE
\begin{equation}
	\label{eq:3d_radial_GPE_equation_phi}
	i\frac{\partial}{\partial\tau}\phi(\rho,\tau)=
	\left\{-\frac{1}{2}\frac{\partial^2}{\partial \rho^2}+ v(\rho,\tau)+ \frac{g N}{\rho^2} \left| \phi(\rho,\tau) \right|^2 \right\} \phi(\rho,\tau)
\end{equation}
for the dimensionless radial wave function $\phi(\rho,\tau)\equiv \sqrt{4\pi}a_{\mathrm{HO}}^{3/2}\,\rho\,\psi(\rho,\tau)$. The interaction constant $g=a_s/a_{\mathrm{HO}}$ and the dimensionless external potential $v(\rho,\tau)$ is given by
\begin{equation}
	\label{eq:Radial_HO_potential}
	v(\rho,\tau) = \frac{1}{2}\left[f(\tau)\right]^2(\rho-\rho_0)^2
\end{equation}
with $\rho_0\equiv r_0/a_{\mathrm{HO}}$.

Moreover, it is worth emphasizing that we consider only the spherically symmetric solution of the 3D GPE (\ref{eq_TB_GPE_full3D}) and the normalization condition for $\phi(\rho,\tau)$ now reads
\begin{equation}
 \label{eq:normalisation_condition_phi}
    \int_{0}^{\infty}d\rho \abs{\phi(\rho,\tau)}^{2} = 1.
\end{equation}

Now we investigate the role of atom-atom interaction in a shell-shaped BEC, prepared as the ground state in the potential $v(\rho)=(\rho-\rho_0)^2/2$, that is $v(\rho,\tau)$, \eqref{eq:Radial_HO_potential}, with $f(\tau)=1$. With the help of the imaginary-time propagation method \cite{auer_fourth-order_2001}, we have solved the time-dependent GPE (\ref{eq:3d_radial_GPE_equation_phi}) numerically and obtained the radial wave function $\phi_0(\rho)$ for the ground state of the BEC. 

This wave function $\phi_0(\rho)$ is normalized according to the condition \eqref{eq:normalisation_condition_phi} and determines the dependence of the kinetic, potential, interaction, and total energy per particles
\begin{eqnarray}
	\label{eq:Energy_Ekin}
	\mathcal{E}_{\mathrm{kin}} &\equiv \int_0^\infty d\rho \phi_0^*(\rho) \left(-\frac{1}{2}\frac{\partial^2}{\partial \rho^2}\phi_0(\rho)\right) =\frac{1}{2}\int_{0}^{\infty}d\rho\Big|\frac{\partial}{\partial \rho}\phi_0(\rho)\Big|^2,\\
	\label{eq:Energy_Epot}
	\mathcal{E}_{\mathrm{pot}} &\equiv \frac{1}{2}\int_0^\infty d\rho (\rho-\rho_0)^2 |\phi_0(\rho)|^2, \\
	\label{eq:Energy_Eint}
	\mathcal{E}_{\mathrm{int}} &\equiv \frac{1}{2} \int_0^\infty d\rho \left(\frac{gN}{\rho^2}\right)|\phi_0(\rho)|^4,\\
	\label{eq:Energy_Etot}
	\mathcal{E}_{\mathrm{GP}} &\equiv \mathcal{E}_{\mathrm{kin}} + \mathcal{E}_{\mathrm{pot}} + \mathcal{E}_{\mathrm{int}},
\end{eqnarray}
on the position $\rho_0$ of the trap minimum and the interaction constant $g$. 

\begin{figure}[t]
	\centering
	\includegraphics{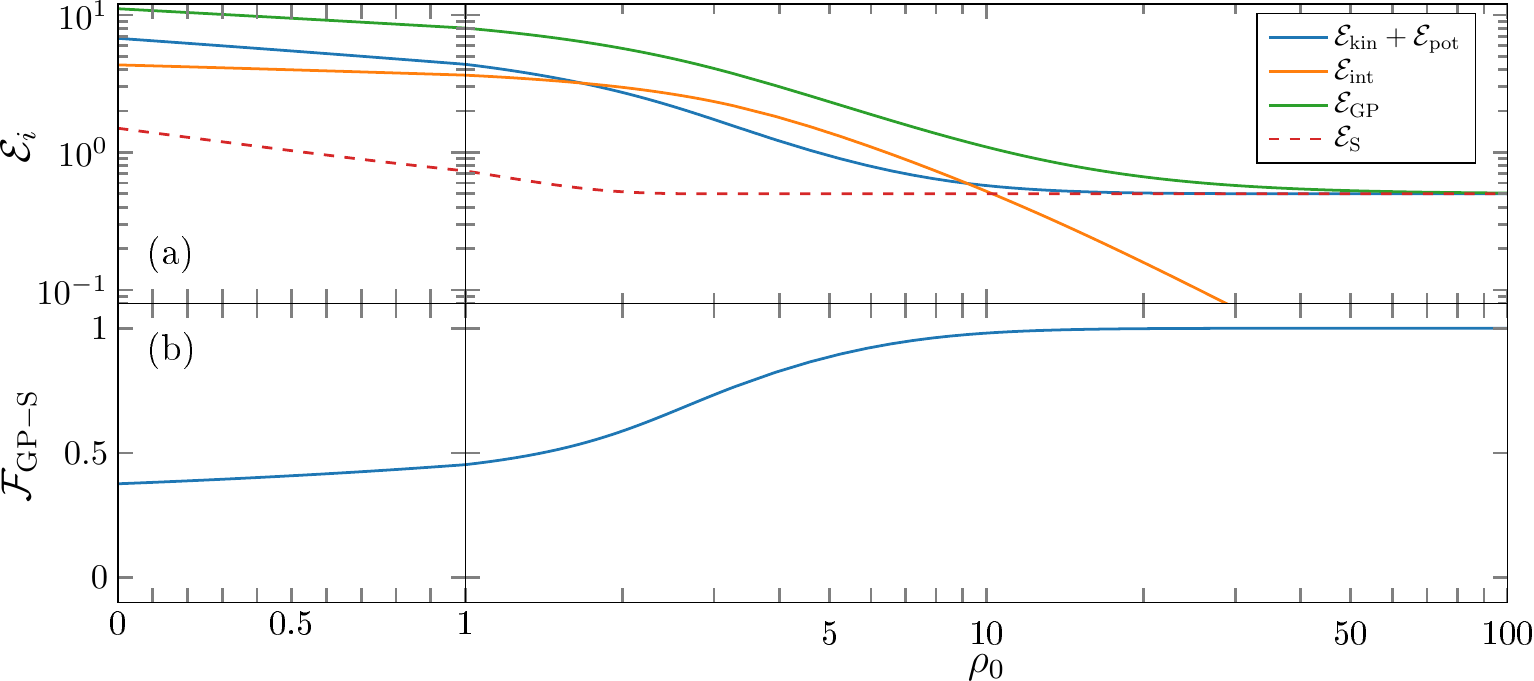}
	\caption{Dependence of the energies per particles ${\mathcal E}_{\mathrm{ kin}}+{\mathcal E}_{\mathrm{pot}}$, ${\mathcal E}_{\mathrm{int}}$, and ${\mathcal E}_{\mathrm{GP}}$, defined by Eqs. (\ref{eq:Energy_Ekin})-(\ref{eq:Energy_Etot}), as well as the overlap ${\mathcal F}_{\mathrm{GP-S}}$, \eqref{eq:Fidelity_GPE}, on the minimum position $\rho_0$ of the potential $(\rho-\rho_0)^2/2$. The solid lines in (a) and (b) are plotted for $g=0.0035$ and $N = 10^5$, whereas the dashed red line corresponds to the total energy per particle ${\mathcal E}_{\mathrm{S}}$ of non-interacting atoms, that is $g=0$. The horizontal axis of both figures has two different scaling: a linear one, for $0\leq \rho_0 \leq 1$, and a logarithmic one, for $1\leq \rho_0 \leq 100$. \label{fig: GPE_Schr_comparison}}
\end{figure}

The energies ${\mathcal E}_{\mathrm{ kin}}+{\mathcal E}_{\mathrm{pot}}$, ${\mathcal E}_{\mathrm{int}}$, and ${\mathcal E}_{\mathrm{GP}}$ are presented by corresponding solid lines in \figref{fig: GPE_Schr_comparison} (a) for different values of $\rho_0$ with $N=10^5$ and $g=a_s/a_{\mathrm{HO}}=0.0035$. This value of $g$ corresponds to $^{87}\mathrm{Rb}$ atoms, the scattering length $a_s=100 a_0$, and the trapping frequency $\omega_0=2\pi\cdot 50\,\mathrm{Hz}$, where $a_0$ is the Bohr radius.  

For small $\rho_0$, $0\leq \rho_0\lesssim 10$, the total energy $\mathcal{E}_{\mathrm{GP}}(\rho_0,g)$, green line, differs significantly from the total energy $\mathcal{E}_{\mathrm{S}}\equiv\mathcal{E}_{\mathrm{GP}}(\rho_0,0)$ per particle of non-interacting atoms, dashed red line, that is the energy of the ground-state in the potential $v(\rho)$, \eqref{eq:Radial_HO_potential}, obtained by solving the Schr\"odinger equation. However, for large $\rho_0$, $\rho_0>10$, the interaction energy $\mathcal{E}_{\mathrm{int}}(\rho_0, g)$, orange line, is substantially reduced and both the total energy $\mathcal{E}_{\mathrm{GP}}(\rho_0, g)$ and the sum ${\mathcal E}_{\mathrm{ kin}}+{\mathcal E}_{\mathrm{pot}}$, blue line, therefore approach $\mathcal{E}_{\mathrm{S}}$.

In addition, we have calculated the overlap 
\begin{equation}
	\label{eq:Fidelity_GPE}
	\mathcal{F}_{\mathrm{GP-S}}\equiv \int_{0}^{\infty} d\rho \phi_{0}^*(\rho;g) \phi_{0}(\rho;0)
\end{equation}
between the radial wave function $\phi_0(\rho;g)$ of the ground state of an interacting BEC and the one $\phi_0(\rho;0)$ of the a noninteracting BEC, $g=0$. For $\rho_0\rightarrow \infty$, the overlap ${\mathcal F}_{\mathrm{GP-S}}$ as the function of $\rho_0$ approaches unity, as displayed in \figref{fig: GPE_Schr_comparison} (b).

Hence, we have shown that for large enough values of $\rho_0$ both the energy and the wave function of the ground state of a BEC in the potential $v(\rho)=(\rho-\rho_0)^2/2$ could be perfectly described by a 3D Schr\"odinger equation rather than a 3D GPE.

\subsection{From 3D to 1D Schrödinger ground-state}
\label{Appendix:3D-1D}

By using a 3D stationary Schrödinger equation, in this section we obtain the energy and the corresponding wave function of the ground state in the potential $v(\rho)=(\rho-\rho_0)^2/2$ for any value of $\rho_0$. In addition, we show that for large $\rho_0$ this state coincides with the one of a 1D harmonic oscillator.

The general solution of the 3D stationary Schrödinger equation
\begin{equation}
	\label{eq:3d_radial_GPE_equation_phi_static}
	-\frac{1}{2}\frac{\partial^2}{\partial \rho^2} \phi(\rho)+ \frac{1}{2} (\rho-\rho_0)^2 \phi(\rho)=E \phi(\rho)
\end{equation}
for the wave function $\phi(\rho)$ with the corresponding energy $E$ is given by the linear superposition  
\begin{equation}
	\phi(\rho) = A D_\nu \left[\sqrt{2} (\rho-\rho_0)\right] + B D_{-1-\nu}\left[i\sqrt{2} (\rho-\rho_0)\right]
\end{equation}
of the parabolic cylindrical functions $D_\nu(z)$ and $D_{-1-\nu}(iz)$ \cite{national_institute_of_standards_and_technology_nist_2010} with $\nu \equiv E-1/2$.

The unknown constants $A$ and $B$ are determined by the boundary conditions
\begin{eqnarray}
	\lim_{\rho \rightarrow 0} \phi(\rho) = 0 \mathrm{~~~and~~~}\lim_{\rho \rightarrow \infty } \phi(\rho) = 0,
\end{eqnarray}
giving rise to the wave function
\begin{eqnarray}
	\label{eq:3DSchroedinger_solution}
	\phi_n(\rho) = N_n(r_0) D_{E_n - \frac{1}{2}} \left[\sqrt{2} (\rho-\rho_0)\right]
\end{eqnarray}
of the bound state ($n=0,1,2,...$) with the energy $E_n$, where we have used the fact that the function $D_{-1-\nu}(iz)$ diverges for $z\rightarrow +\infty$ \cite{national_institute_of_standards_and_technology_nist_2010}. In addition, the normalisation constant $N_n(\rho_0)$ reads
\begin{equation}
    \label{eq:3DSchroedinger_solution_normalization}
	N_n(\rho_0) = \left(4\pi\int_{0}^{\infty} d\rho \left\{D_{E_n - \frac{1}{2}}\left[\sqrt{2} (\rho-\rho_0)\right]\right\}^2\right)^{-\frac{1}{2}},
\end{equation}
whereas the energy $E_n$ is defined as a solution (with $E_n\geq 0$) of the transcendental equation
\begin{equation}
    D_{E_n - \frac{1}{2}} \left(-\sqrt{2} \rho_0\right) = 0
	\label{eq:3DSchroedinger_solution_energy}
\end{equation}
for any given $\rho_0 \geq 0$.

Here we are only interested in finding the dependence of the energy $E_0$ of the ground state on $\rho_0$. This is obtained by solving \eqref{eq:3DSchroedinger_solution_energy}. However, this equation is quite complicated and we therefore derive analytical formulas for $E_0(\rho_0)$ in two limiting cases of (i) small and (ii) large values of $\rho$.  

In the case of a small shell radius, $\rho_0 \rightarrow 0$, we insert the Taylor series for the function $D_{E_n - \frac{1}{2}} \left(-\sqrt{2} \rho_0\right)$ at its small argument into \eqref{eq:3DSchroedinger_solution_energy} and obtain the asymptotic expansion  
\begin{equation}
    \label{eq:E0-small-r0}
	E_0(\rho_0) \simeq \frac{3}{2} - \frac{2}{\sqrt{\pi}}\rho_0 + \frac{4}{\pi} \left[1-\ln(2)\right]\rho_0^2
\end{equation}
for the ground-state energy $E_0(\rho_0)$ as $\rho_0\rightarrow 0$, where we have omitted all terms of smaller orders.
It means that the ground-state energy $E_0(\rho_0)$ approaches the one of a 3D spherically symmetric harmonic oscillator, $E_0(0)=3/2$. In this limit, the corresponding ground-state wave function
\begin{equation}
	\phi_0(\rho) \simeq N_0(0) D_1\left(\sqrt{2} \rho\right) =\frac{\rho}{\pi^{3/4}}\exp\left(-\frac{\rho^2}{2}\right) 
\end{equation}
also coincides with the one of the ground state of a 3D spherically symmetric harmonic oscillator. Here we have used the fact that $D_1(z)=z\exp(-z^2/4)$ \cite{national_institute_of_standards_and_technology_nist_2010} and calculated the normalization constant $N_0$, \eqref{eq:3DSchroedinger_solution_normalization}, for $\rho_0=0$.   

In the opposite case of a shell with large radius, $\rho_0 \rightarrow \infty$, we insert the Taylor series for the function $D_{E_n - \frac{1}{2}} \left(-\sqrt{2} \rho_0\right)$, now at its large negative argument, into \eqref{eq:3DSchroedinger_solution_energy} and derive the asymptotic expansion  
\begin{equation}
    \label{eq:E0-large-r0}
	E_0(\rho_0) \simeq \frac{1}{2} + \frac{\rho_0}{\sqrt{\pi}} \mathrm{e}^{-\rho_0^2}
\end{equation}
for the ground-state energy $E_0(\rho_0)$ as $\rho_0\rightarrow \infty$.

Since $E_0(\rho_0)=1/2$ as $\rho_0\rightarrow \infty$, the wave function of the ground state
\begin{equation}
    \label{eq:App1d-wave-function}
	\phi_0(\rho) \simeq \Phi_{\mathrm{1D}}(\rho)=N_0(\infty) D_0\left[\sqrt{2} \left(\rho-\rho_0\right)\right] = \frac{1}{2\pi^{3/4}}\exp\left[-\frac{(\rho-\rho_0)^2}{2}\right]
\end{equation}
approaches the one of the ground state of a 1D harmonic oscillator with the minimum located at $\rho_0$. In order to derive this result, we have used the relation $D_0(z)=\exp(-z^2/4)$ \cite{national_institute_of_standards_and_technology_nist_2010} and calculated again the corresponding normalization factor $N_0(\infty)$, \eqref{eq:3DSchroedinger_solution_normalization}, for $\rho_0\rightarrow \infty$.

\begin{figure}[h]
	\centering
	\includegraphics{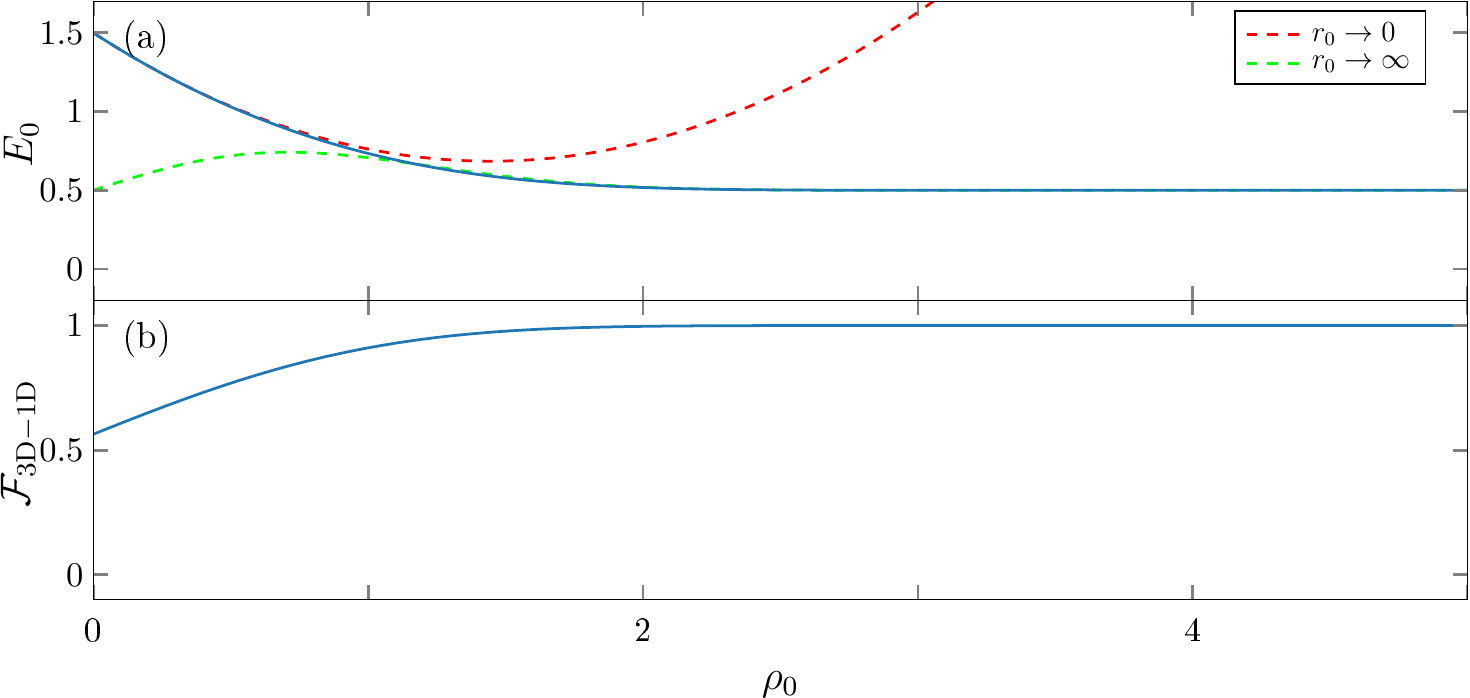}
	\caption{Dependence of the ground-state energy $E_0$, defined by \eqref{eq:3DSchroedinger_solution_energy}, of the 3D Schrödinger equation, \eqref{eq:3d_radial_GPE_equation_phi_static}, and the overlap $\mathcal{F}_{\mathrm{3D-1D}}$, \eqref{eq:appFidelity1D}, on the shift $\rho_0$ of the 3D spherically symmetric harmonic oscillator. The dashed red and green lines describe two asymptotic expansions for small, \eqref{eq:E0-small-r0}, and large, \eqref{eq:E0-large-r0}, values of $\rho_0$, accordingly. \label{fig:Schroedinger_comparison}}
\end{figure}

To find the ground-state energy $E_0$ for any value of $\rho_0$, we solve \eqref{eq:3DSchroedinger_solution_energy} numerically and present this result in \figref{fig:Schroedinger_comparison} (a) by the solid green line. For small and large values of $\rho_0$, this line perfectly reproduces the asymptotic behaviors given by \eqref{eq:E0-small-r0} and \eqref{eq:E0-large-r0}, accordingly. 

In addition, we have calculated the overlap 
\begin{equation}
	\label{eq:appFidelity1D}
	\mathcal{F}_{\mathrm{3D-1D}} = \int_{0}^{\infty} d\rho \phi_{0}^*(\rho) \Phi_{\mathrm{1D}}(\rho)
\end{equation}
between the 3D wave function $\phi_0(\rho)$, \eqref{eq:3DSchroedinger_solution}, of the ground state and the corresponding 1D wave function $\Phi_{\mathrm{1D}}(\rho)$, \eqref{eq:App1d-wave-function},
and present its dependence on $\rho_0$ in \figref{fig:Schroedinger_comparison} (b). It shows in the clearest way that the ground state of the spherically symmetric harmonic oscillator with large shift $\rho_0$ is well described by the one of the shifted 1D harmonic potential. 

As a result, according to the consideration presented in \ref{Appendix:role_of_atom_atom_interaction}, our system is already in the regime of large $\rho_0$ and therefore we can model the trapping potential by the 1D shifted harmonic oscillator and study collimation of a shell-shaped BEC.

\section{The effective 1D analytical model}
\label{Appendix:effective_1D_model}

In this Appendix we consider the dynamics of the wave function, being initially prepared in a Gaussian shape, in the one-dimensional harmonic oscillator with a time-dependent frequency. 
   
\subsection{Quantum dynamics of the 1D harmonic oscillator with time-dependent frequency}

The solution of the one-dimensional Schr\"odinger equation
\begin{equation}
 \label{eq:app_1d_Schroedinger_equation}
    i\frac{\partial}{\partial\tau}\varphi(\xi,\tau)=
    \left\{-\frac{1}{2}\frac{\partial^2}{\partial \xi^2}+\frac{1}{2}[f(\tau)]^2(\xi-\rho_0)^2\right\}\varphi(\xi,\tau)
\end{equation}
for the wave function $\varphi(\xi,\tau)$, with $\xi\equiv x/a_\mathrm{HO}$ being the dimensionless coordinate, $\xi\in (-\infty,\infty)$, governed by a harmonic oscillator with a time-dependent dimensionless frequency $f(\tau)$ is given by 
\begin{equation}
 \label{eq:app_1d_general_solution}
    \varphi(\xi,\tau)=\int_{-\infty}^{+\infty}dy\,G(\xi,\tau;\eta,0)\varphi(\eta,0),
\end{equation}
where $\varphi(\xi,0)$ is the initial wave function. 

The Green function
\begin{eqnarray}
 \label{eq:app_1d_Green_function}
    G(\xi,\tau;\eta,0) & = &\frac{1}{\sqrt{2i\pi\lambda(\tau)\lambda(0)\sin\left[\Phi(\tau)-\Phi(0)\right]}}\nonumber\\
    &\times & \exp\left\{i\frac{\dot{\lambda}(\tau)}{2\lambda(\tau)}(\xi-\rho_0)^2-i\frac{\dot{\lambda}(0)}{2\lambda(0)}(\eta-\rho_0)^2\right\}\nonumber\\
     &\times & \exp\left\{i\left[\frac{(\xi-\rho_0)^2}{2[\lambda(\tau)]^2}+\frac{(\eta-\rho_0)^2}{2[\lambda(0)]^2}\right]\cot\left[\Phi(\tau)-\Phi(0)\right]\right\}\nonumber\\
     & \times & \exp\left\{-i\frac{(\xi-\rho_0)(\eta-\rho_0)}{\lambda(\tau)\lambda(0)\sin\left[\Phi(\tau)-\Phi(0)\right]}\right\}
\end{eqnarray}
is determined by the time-dependent functions 
\begin{equation}
 \label{eq:app_1d_general_solution_T}
    \Phi(\tau)=\int_{0}^{\tau}\frac{d\tau'}{[\lambda(\tau')]^2}
\end{equation}
and $\lambda(\tau)$ with its derivative $\dot{\lambda}(\tau)\equiv d\lambda(\tau)/d\tau$.

The function $\lambda(\tau)$ is the solution of the Ermakov equation
\begin{equation}
 \label{eq:app_Ermakov_equation}
    \frac{d^2}{d\tau^2}\lambda+[f(\tau)]^2\lambda=\frac{1}{\lambda^3}
\end{equation}
with the initial condition $\lambda(0)=1$ and $\dot{\lambda}(0)=0$. 

Taking the initial wave function $\varphi(\xi,0)$ in the form of the normalized wave function
\begin{equation}
 \label{eq:app_initial_wave_function}
    \varphi(\xi,0)=\frac{1}{\pi^{1/4}}\exp\left(-\frac{(\xi-\rho_0)^2}{2}\right)
\end{equation}
of the ground state of the harmonic oscillator \eqref{eq:app_1d_Schroedinger_equation} with $f=1$, we perform the integration in \eqref{eq:app_1d_general_solution} and arrive at
\begin{equation}
 \label{eq:initial_wave_function_result}
    \varphi(\xi,\tau)=\left[\frac{1}{\sqrt{\pi}\lambda(\tau)}\right]^{1/2}\exp\left[-\frac{(1-i\lambda\dot{\lambda})}{2\lambda^2}(\xi-\rho_0)^2-\frac{i}{2}\Phi(\tau)\right].
\end{equation}
Here we have used the facts that $\Phi(0)=0$ and $\lambda(0)=1$.

\subsection{Solution of the Ermakov equation}

The solution of the general form of the Ermakov equation
\begin{equation}
 \label{eq:Ermakov equation_general}
    \frac{d^2}{d\tau^2}\lambda+[f(\tau)]^2\lambda=\frac{a}{\lambda^3},
\end{equation}
with the initial conditions $\lambda(0)=\alpha$ and $\dot{\lambda}(0)=\beta$, where $a$, $\alpha$ and $\beta$ are some constants, is given by
\begin{equation}
 \label{eq:Ermakov_equation_general_solution}
    \lambda(\tau)=\left\{\left[\alpha\Lambda_1(\tau)+\beta\Lambda_2(\tau)\right]^2+\frac{a}{\alpha^2}\left[\Lambda_2(\tau)\right]^2\right\}^{\frac{1}{2}}.
\end{equation}
Here $\Lambda_1(\tau)$ and $\Lambda_2(\tau)$ are two linearly independent solutions of the corresponding second-order linear differential equation 
\begin{equation}
 \label{eq:app_equation_for_xi}
    \frac{d^2}{d\tau^2}\Lambda+[f(\tau)]^2\Lambda=0
\end{equation}
with the initial conditions: $\Lambda_1(0)=1$ and $\dot{\Lambda}_1(0)=0$, or $\Lambda_2(0)=0$ and $\dot{\Lambda}_2(0)=1$.

In the case of the generalized scheme of DKC the profile $f(\tau)$ of the trapping frequency reads
\begin{equation}
 \label{eq:app_general_profile}
    f_{\mathrm{gDKC}}(\tau)=
    \begin{cases}
    f_1,\;\;\; 0<\tau\leq\tau_d\\
    f_2, \;\;\; \tau_d<\tau\leq\tau_d+\tau_k\\
    0, \;\;\; \tau_d+\tau_k\leq \tau
    \end{cases}
\end{equation}
with the constants $f_1$ and $f_2$, where $\tau_d$ and $\tau_k$ are the dimensionless time delay and the kick duration, respectively.

By taking $a=1$, $\alpha=1$, and $\beta=0$ in \eqref{eq:Ermakov_equation_general_solution} and solving \eqref{eq:app_equation_for_xi} with $f(\tau)$, \eqref{eq:app_general_profile}, we obtain the solution of \eqref{eq:app_Ermakov_equation} in the form 
\begin{equation}
 \label{eq:app_Ermakov_equation_solution}
    \lambda_{\mathrm{gDKC}}(\tau)=\sqrt{\left[\Lambda_1^{\mathrm{(gDKC)}}(\tau)\right]^2+\left[\Lambda_2^{\mathrm{(gDKC)}}(\tau)\right]^2}, 
\end{equation}
with $\Lambda_1^{\mathrm{(gDKC)}}(\tau)$ and $\Lambda_2^{\mathrm{(gDKC)}}(\tau)$ given by 
\begin{equation}
 \label{eq:lambda1_gdkc}
    \Lambda_1^{\mathrm{(gDKC)}}(\tau)=
    \begin{cases}
    \cos(f_1\tau),\; 0<\tau\leq\tau_d\\
    \cos(f_1\tau_d)\cos[f_2(\tau-\tau_d)]-(f_1/f_2)\sin(f_1\tau_d)\sin[f_2(\tau-\tau_d)], \; \tau_d<\tau\leq\tau_d+\tau_k\\
    \cos(f_1\tau_d)\cos(f_2\tau_k)-(f_1/f_2)\sin(f_1\tau_d)\sin(f_2\tau_k)-\\
    [f_2\cos(f_1\tau_d)\sin(f_2\tau_k)+f_1\sin(f_1\tau_d)\cos(f_2\tau_k)](\tau-\tau_d-\tau_k), \; \tau_d+\tau_k\leq \tau
    \end{cases}
\end{equation}
and 
\begin{equation}
 \label{eq:lambda2_gdkc}
    \Lambda_2^{\mathrm{(gDKC)}}(\tau)=\frac{1}{f_1}
    \begin{cases}
    \sin(f_1\tau),\; 0<\tau\leq\tau_d\\
    \sin(f_1\tau_d)\cos[f_2(\tau-\tau_d)]+(f_1/f_2)\cos(f_1\tau_d)\sin[f_2(\tau-\tau_d)], \; \tau_d<\tau\leq\tau_d+\tau_k\\
    \sin(f_1\tau_d)\cos(f_2\tau_k)+(f_1/f_2)\cos(f_1\tau_d)\sin(f_2\tau_k)-\\
    [f_2\sin(f_1\tau_d)\sin(f_2\tau_k)-f_1\cos(f_1\tau_d)\cos(f_2\tau_k)](\tau-\tau_d-\tau_k), \; \tau_d+\tau_k\leq \tau,
    \end{cases}
\end{equation}
respectively. 

In the case of the rf-dressing scheme we model the profile $f_{\mathrm{RF}}(\tau)$ of the trapping frequency by
\begin{equation}
 \label{eq:app_general_profile_RF}
    f_{\mathrm{RF}}(\tau)=
    \begin{cases}
    1-(1-f_1)(\tau/\tau_r),\;\;\; 0<\tau\leq\tau_r\\
    f_1, \;\;\;\;\; \tau_r<\tau\leq\tau_r+\tau_k\\
    0, \;\;\;\;\; \tau_r+\tau_k\leq \tau,
    \end{cases}
\end{equation}
where $f_1$ is the constant, $0<f_1<1$, $\tau_r$ and $\tau_k$ are the dimensionless ramping time and the kick duration, respectively.

The solutions $\Lambda_1^{\mathrm{(RF)}}(\tau)$ and $\Lambda_2^{\mathrm{(RF)}}(\tau)$ of \eqref{eq:app_equation_for_xi} corresponding to $f_{\mathrm{RF}}(\tau)$ read 
\begin{equation}
 \label{eq:lambda1_RF}
    \Lambda_1^{\mathrm{(RF)}}(\tau)=
    \begin{cases}
    (\pi/2)z_0^{3/4}z^{1/4}\left[Y_{-3/4}(z_0)J_{1/4}(z)-J_{-3/4}(z_0)Y_{1/4}(z)\right],\; 0<\tau\leq\tau_r\\
    \Lambda_1^{\mathrm{(RF)}}(\tau_r)\cos[f_1(\tau-\tau_r)]+[\dot{\Lambda}_1^{\mathrm{(RF)}}(\tau_r)/f_1]\sin[f_1(\tau-\tau_r)], \; \tau_r<\tau\leq\tau_r+\tau_k\\
    \Lambda_1^{\mathrm{(RF)}}(\tau_r)\cos(f_1\tau_k)+[\dot{\Lambda}_1^{\mathrm{(RF)}}(\tau_r)/f_1]\sin(f_1\tau_k)-\\
    [\Lambda_1^{\mathrm{(RF)}}(\tau_r)f_1\sin(f_1\tau_k)-\dot{\Lambda}_1^{\mathrm{(RF)}}(\tau_r)\cos(f_1\tau_k)](\tau-\tau_r-\tau_k), \; \tau_r+\tau_k\leq \tau
    \end{cases}
\end{equation}
and 
\begin{equation}
 \label{eq:lambda2_RF}
    \Lambda_2^{\mathrm{(RF)}}(\tau)=
    \begin{cases}
    (\pi/2)z_0^{3/4}z^{1/4}\left[Y_{1/4}(z_0)J_{1/4}(z)-J_{1/4}(z_0)Y_{1/4}(z)\right],\; 0<\tau\leq\tau_r\\
    \Lambda_2^{\mathrm{(RF)}}(\tau_r)\cos[f_1(\tau-\tau_r)]+[\dot{\Lambda}_2^{\mathrm{(RF)}}(\tau_r)/f_1]\sin[f_1(\tau-\tau_r)], \; \tau_r<\tau\leq\tau_r+\tau_k\\
    \Lambda_2^{\mathrm{(RF)}}(\tau_r)\cos(f_1\tau_k)+[\dot{\Lambda}_2^{\mathrm{(RF)}}(\tau_r)/f_1]\sin(f_1\tau_k)-\\
    [\Lambda_2^{\mathrm{(RF)}}(\tau_r)f_1\sin(f_1\tau_k)-\dot{\Lambda}_2^{\mathrm{(RF)}}(\tau_r)\cos(f_1\tau_k)](\tau-\tau_r-\tau_k), \; \tau_r+\tau_k\leq \tau,
    \end{cases}
\end{equation}
respectively. Here $J_\nu(z)$ and $Y_\nu(z)$ are the Bessel functions of the first and second kind \cite{national_institute_of_standards_and_technology_nist_2010}. Moreover, we have introduced the function
\begin{equation}
    z=z(\tau)=\frac{\tau_r}{2(1-f_1)}\left[1-(1-f_1)\frac{\tau}{\tau_r}\right]^2 
\end{equation}
as well as notations $z_0=z(0)=\tau_r/[2(1-f_1)]$ and $z_1=z(\tau_r)=\tau_rf_1^2/[2(1-f_1)]$. The constants $\Lambda_{1,2}^{\mathrm{(RF)}}(\tau_r)$ and $\dot{\Lambda}_{1,2}^{\mathrm{(RF)}}(\tau_r)$ are given by
\begin{equation}
   \begin{split}
    \Lambda_1^{\mathrm{(RF)}}(\tau_r)&=\frac{\pi}{2}z_0^{3/4}z_1^{1/4}\left[Y_{-3/4}(z_0)J_{1/4}(z_1)-J_{-3/4}(z_0)Y_{1/4}(z_1)\right]\\
    \dot{\Lambda}_1^{\mathrm{(RF)}}(\tau_r)&= -\frac{\pi}{2}z_0^{3/4}z_1^{1/4}f_1\left[Y_{-3/4}(z_0)J_{-3/4}(z_1)-J_{-3/4}(z_0)Y_{-3/4}(z_1)\right]
   \end{split}
\end{equation}
and 
\begin{equation}
   \begin{split}
    \Lambda_2^{\mathrm{(RF)}}(\tau_r)&=\frac{\pi}{2}z_0^{3/4}z_1^{1/4}\left[Y_{1/4}(z_0)J_{1/4}(z_1)-J_{1/4}(z_0)Y_{1/4}(z_1)\right]\\
    \dot{\Lambda}_2^{\mathrm{(RF)}}(\tau_r)&= -\frac{\pi}{2}z_0^{3/4}z_1^{1/4}f_1\left[Y_{1/4}(z_0)J_{-3/4}(z_1)-J_{1/4}(z_0)Y_{-3/4}(z_1)\right].
   \end{split}
\end{equation}

\section*{References}
\bibliographystyle{ieeetr}
\bibliography{bibliographymaster.bib}

\begin{thebibliography}{10}

\bibitem{Carollo.2022}
R.~A. Carollo, D.~C. Aveline, B.~Rhyno, S.~Vishveshwara, C.~Lannert, J.~D.
  Murphree, E.~R. Elliott, J.~R. Williams, R.~J. Thompson, and N.~Lundblad,
  ``{Observation of ultracold atomic bubbles in orbital microgravity},'' {\em
  Nature}, vol.~606, no.~7913, pp.~281--286, 2022.

\bibitem{Jia.2022}
F.~Jia, Z.~Huang, L.~Qiu, R.~Zhou, Y.~Yan, and D.~Wang, ``Expansion dynamics of
  a shell-shaped {Bose-Einstein} condensate.'' arXiv:2208.01360, 2022.

\bibitem{Turner.2010}
A.~M. Turner, V.~Vitelli, and D.~R. Nelson, ``{Vortices on curved surfaces},''
  {\em Rev. Mod. Phys.}, vol.~82, no.~2, pp.~1301--1348, 2010.

\bibitem{Padavic.2020}
K.~Padavi{\'c}, K.~Sun, C.~Lannert, and S.~Vishveshwara, ``{Vortex-antivortex
  physics in shell-shaped Bose-Einstein condensates},'' {\em Phys. Rev. A},
  vol.~102, no.~4, p.~043305, 2020.

\bibitem{Bereta.2021}
S.~J. Bereta, M.~A. Caracanhas, and A.~L. Fetter, ``{Superfluid vortex dynamics
  on a spherical film},'' {\em Phys. Rev. A}, vol.~103, no.~5, p.~053306, 2021.

\bibitem{Kosterlitz.2016}
J.~M. Kosterlitz, ``{Kosterlitz-Thouless physics: a review of key issues},''
  {\em Rep. Prog. Phys.}, vol.~79, no.~2, p.~026001, 2016.

\bibitem{Tononi.2021}
A.~Tononi, A.~Pelster, and L.~Salasnich, ``{Topological superfluid transition
  in bubble-trapped condensates},'' {\em Phys. Rev. Research}, vol.~4, no.~1,
  p.~013122, 2022.

\bibitem{Padavic.2017}
K.~Padavi{\'c}, K.~Sun, C.~Lannert, and S.~Vishveshwara, ``{Physics of hollow
  Bose-Einstein condensates},'' {\em EPL}, vol.~120, no.~2, p.~20004, 2017.

\bibitem{Sun.2018}
K.~Sun, K.~Padavi{\'c}, F.~Yang, S.~Vishveshwara, and C.~Lannert, ``{Static and
  dynamic properties of shell-shaped condensates},'' {\em Phys. Rev. A},
  vol.~98, no.~1, p.~013609, 2018.

\bibitem{Perez-Rios.2020}
J.~P. R{\'{\i}}os, {\em An Introduction to Cold and Ultracold Chemistry}.
\newblock Cham: Springer, 2020.

\bibitem{Naidon_2017}
P.~Naidon and S.~Endo, ``Efimov physics: a review,'' {\em Reports on Progress
  in Physics}, vol.~80, p.~056001, mar 2017.

\bibitem{Blume_2012}
D.~Blume, ``Few-body physics with ultracold atomic and molecular systems in
  traps,'' {\em Reports on Progress in Physics}, vol.~75, p.~046401, mar 2012.

\bibitem{Olshani_PRL1998}
M.~Olshanii, ``Atomic scattering in the presence of an external confinement and
  a gas of impenetrable bosons,'' {\em Phys. Rev. Lett.}, vol.~81,
  pp.~938--941, Aug 1998.

\bibitem{DUNJKO2011461}
V.~Dunjko, M.~G. Moore, T.~Bergeman, and M.~Olshanii, ``Chapter 10 -
  confinement-induced resonances,'' in {\em Advances in Atomic, Molecular, and
  Optical Physics} (E.~Arimondo, P.~Berman, and C.~Lin, eds.), vol.~60 of {\em
  Advances In Atomic, Molecular, and Optical Physics}, pp.~461--510, Academic
  Press, 2011.

\bibitem{Zobay.2001}
O.~Zobay and B.~M. Garraway, ``{Two-dimensional atom trapping in field-induced
  adiabatic potentials},'' {\em Phys. Rev. Lett.}, vol.~86, no.~7,
  pp.~1195--1198, 2001.

\bibitem{Zobay.2004}
O.~Zobay and B.~M. Garraway, ``{Atom trapping and two-dimensional Bose-Einstein
  condensates in field-induced adiabatic potentials},'' {\em Phys. Rev. A},
  vol.~69, no.~2, p.~023605, 2004.

\bibitem{Garraway.2016}
B.~M. Garraway and H.~Perrin, ``{Recent developments in trapping and
  manipulation of atoms with adiabatic potentials},'' {\em J. Phys. B: At. Mol.
  Opt. Phys.}, vol.~49, no.~17, p.~172001, 2016.

\bibitem{Perrin.2017}
H.~Perrin and B.~M. Garraway, ``{Trapping Atoms With Radio Frequency Adiabatic
  Potentials},'' in {\em {Advances in Atomic, Molecular, and Optical Physics}}
  (S.~F. Yelin, E.~Arimondo, and C.~C. Lin, eds.), vol.~66, pp.~181--262, Saint
  Louis: {Elsevier Science}, 2017.

\bibitem{Lundblad.2019}
N.~Lundblad, R.~A. Carollo, C.~Lannert, M.~J. Gold, X.~Jiang, D.~Paseltiner,
  N.~Sergay, and D.~C. Aveline, ``{Shell potentials for microgravity
  Bose-Einstein condensates},'' {\em npj Microgravity}, vol.~5, p.~30, 2019.

\bibitem{Wolf.2022}
A.~Wolf, P.~Boegel, M.~Meister, A.~Bala{\v{z}}, N.~Gaaloul, and M.~A. Efremov,
  ``{Shell-shaped Bose-Einstein condensates based on dual-species mixtures},''
  {\em Phys. Rev. A}, vol.~106, no.~1, 2022.

\bibitem{Meister2022}
M.~Meister and A.~Roura, ``Efficient matter-wave lensing of ultracold atomic
  mixtures,'' 2022.
\newblock arXiv:2207.07045.

\bibitem{Timmermans.1999}
E.~Timmermans, P.~Tommasini, M.~Hussein, and A.~Kerman, ``{Feshbach resonances
  in atomic Bose--Einstein condensates},'' {\em Phys. Rep.}, vol.~315, no.~1-3,
  pp.~199--230, 1999.

\bibitem{Chin.2010}
C.~Chin, R.~Grimm, P.~Julienne, and E.~Tiesinga, ``{Feshbach resonances in
  ultracold gases},'' {\em Rev. Mod. Phys.}, vol.~82, no.~2, pp.~1225--1286,
  2010.

\bibitem{Lannert.2007}
C.~Lannert, T.-C. Wei, and S.~Vishveshwara, ``{Dynamics of condensate shells:
  Collective modes and expansion},'' {\em Phys. Rev. A}, vol.~75, no.~1,
  p.~013611, 2007.

\bibitem{Tononi.2020}
A.~Tononi, F.~Cinti, and L.~Salasnich, ``{Quantum Bubbles in Microgravity},''
  {\em Phys. Rev. Lett.}, vol.~125, no.~1, p.~010402, 2020.

\bibitem{Chu1986}
S.~Chu, J.~E. Bjorkholm, A.~Ashkin, J.~P. Gordon, and L.~W. Hollberg,
  ``Proposal for optically cooling atoms to temperatures of the order of
  $10^{-6}$ {K},'' {\em Opt. Lett.}, vol.~11, pp.~73--75, Feb 1986.

\bibitem{Ammann1997}
H.~Ammann and N.~Christensen, ``Delta kick cooling: a new method for cooling
  atoms,'' {\em Phys. Rev. Lett.}, vol.~78, pp.~2088--2091, Mar 1997.

\bibitem{Kovachy2015}
T.~Kovachy, J.~M. Hogan, A.~Sugarbaker, S.~M. Dickerson, C.~A. Donnelly,
  C.~Overstreet, and M.~A. Kasevich, ``Matter wave lensing to picokelvin
  temperatures,'' {\em Phys. Rev. Lett.}, vol.~114, p.~143004, 2015.

\bibitem{Pandey_2021}
S.~Pandey, H.~Mas, G.~Vasilakis, and W.~von Klitzing, ``Atomtronic matter-wave
  lensing,'' {\em Physical Review Letters}, vol.~126, p.~170402, apr 2021.

\bibitem{Deppner_2021}
C.~Deppner, W.~Herr, M.~Cornelius, P.~Stromberger, T.~Sternke, C.~Grzeschik,
  A.~Grote, J.~Rudolph, S.~Herrmann, M.~Krutzik, A.~Wenzlawski, R.~Corgier,
  E.~Charron, D.~Gu{\'{e}}ry-Odelin, N.~Gaaloul, C.~Lämmerzahl, A.~Peters,
  P.~Windpassinger, and E.~M. Rasel, ``Collective-mode enhanced matter-wave
  optics,'' {\em Phys. Rev. Lett.}, vol.~127, no.~10, p.~100401, 2021.

\bibitem{auer_fourth-order_2001}
J.~Auer, E.~Krotscheck, and S.~A. Chin, ``A fourth-order real-space algorithm
  for solving local {Schrödinger} equations,'' {\em J. Chem. Phys.}, vol.~115,
  pp.~6841--6846, 2001.

\bibitem{national_institute_of_standards_and_technology_nist_2010}
F.~W.~J. Olver, D.~W. Lozier, R.~F. Boisvert, and C.~W. Clark, eds., {\em
  {NIST} handbook of mathematical functions}.
\newblock Cambridge New York Melbourne: Cambridge University Press, 2010.

\end{thebibliography}

\end{document}